\documentclass[aps,prd,onecolumn,notitlepage,nofootinbib,showpacs,preprintnumbers,superscriptaddress,groupedaddress]{revtex4-1}
\usepackage[T1]{fontenc}
\usepackage[utf8]{inputenc}
\usepackage{bm}
\usepackage{multirow}
\usepackage{amsmath}
\usepackage{amssymb}
\usepackage{esint}
\usepackage{color}
\usepackage{array}
\usepackage{float}
\usepackage{graphicx}
\PassOptionsToPackage{normalem}{ulem}
\usepackage{ulem}
\usepackage{hyperref}

\begin{document}
	
	\title{Higher derivative scalar-tensor theory and spatially covariant gravity: the correspondence}
	
	\author{Xian Gao}%
	\email[Email: ]{gaoxian@mail.sysu.edu.cn}
	\affiliation{%
		School of Physics and Astronomy, Sun Yat-sen University, Guangzhou 510275, China}
	
	\author{Yu-Min Hu}%
	\affiliation{%
		School of Physics and Astronomy, Sun Yat-sen University, Guangzhou 510275, China}
	
	\date{April 16, 2020}
	
	\begin{abstract}
		We investigate the correspondence between generally covariant higher derivative scalar-tensor theory and spatially covariant gravity theory. The building blocks are the scalar field and spacetime curvature tensor together with their generally covariant derivatives for the former, and the spatially covariant geometric quantities together with their spatially covariant derivatives for the later. In the case of a single scalar degree of freedom, they are transformed to each other by gauge fixing and recovering procedures, of which we give the explicit expressions.  We make a systematic classification of all the scalar monomials in the spatially covariant gravity according to the total number of derivatives up to $d=4$, and their correspondence to the scalar-tensor monomials. We discusse the possibility of using spatially covariant monomials to generate ghostfree higher derivative scalar-tensor theories. We also derive the covariant 3+1 decomposition without fixing any specific coordinate, which will be useful when performing a covariant Hamiltonian analysis.
	\end{abstract}
	
	\maketitle

\section{Introduction}

As one of the main theories of modified gravity, the scalar-tensor theory, which introduces additional scalar degree(s) of freedom other than the two tensor degrees of the General Relativity (GR), was extensively studied in the past few decades.
In particular, introducing higher derivatives of the scalar field without the Ostrogradsky ghost(s) \cite{Woodard:2015zca} has attracted much attention. 
The representative achievements are the Horndeski theory \cite{Horndeski:1974wa,Nicolis:2008in,Deffayet:2011gz,Kobayashi:2011nu} as well as the  degenerate higher-order derivative theory \cite{Gleyzes:2014dya,Gleyzes:2014qga,Langlois:2015cwa,Motohashi:2016ftl} (see \cite{Langlois:2018dxi,Kobayashi:2019hrl} for reviews). 
These theories contain derivatives of the scalar field up to the second order.
One may wonder if we can go beyond the second order and consider derivatives of the third order or even higher\footnote{Non-polynomial derivative theories that are infinite order in derivatives have also been studied, see (e.g.,) \cite{Buoninfante:2018lnh,Buoninfante:2020ctr} and references therein.}.

The motivation is of two-fold.
On the theoretical side, derivatives of the third and higher order are of the same importance as high curvature terms \cite{Gao:2020juc}.
Generally, high curvature gravity has ghosts due to the higher derivatives.
Nevertheless, the Chern-Simons gravity \cite{Lue:1998mq,Jackiw:2003pm} and the recently studied ghostfree quadratic gravity \cite{Deruelle:2012xv,Crisostomi:2017ugk} show the possibility of the existence of ghostfree high curvature gravity, which also indicates the existence of ghostfree scalar-tensor theories with derivatives beyond the second order.
On the phenomenological side, the parameter space of the scalar-tensor theory with only the second order derivatives (e.g. the Horndeski theory) are highly restricted, in particular, after taking into account the constraint of the propagation speed of the gravitational waves \cite{Creminelli:2017sry,Sakstein:2017xjx,Ezquiaga:2017ekz,Baker:2017hug,Amendola:2017orw,Langlois:2017dyl} (see \cite{Ezquiaga:2018btd} for a review). 
One may wonder if scalar-tensor theories with even higher order derivatives may pass these observational tests, and supply us a broader playground.

An immediate question is how to construct scalar-tensor theories with the third or even higher order derivatives without ghost(s)?
A generic approach is to built the theory in a straightforward manner by finely tuning the structure of the higher derivatives such that they are degenerate. This possibility has been discussed in \cite{Motohashi:2017eya,Motohashi:2018pxg} in the case of point particles.
Although the generalization to the case of field theory can be done in principle, this approach has already shown its complexity even in the case with only second order derivatives.

An alternative approach to the scalar-tensor theory is to view the ``scalar'' as an effective degree of freedom in the theory instead of a scalar field that arises explicitly.
In fact, the existence of the scalar field with non-vanishing vacuum expectation value breaks the general covariance. Conversely, a gravity theory with broken general covariance naturally possesses more degree(s) of freedom than those of GR.
Theory that is most extensively studied along this direction is the spatially covariant gravity, in which the space-dependent time reparametrization symmetry is broken due to the existence of a scalar field with timelike gradient.
As a result, the residual symmetry is the 3 dimensional spatial symmetry on the hypersurfaces of constant scalar field.
When being written in coordinates that are adapted with this foliation structure, the theory appears to be a pure metric theory with only spatial symmetry. In this sense, we may dub it as spatially covariant gravity.
The well-studied effective field theory of inflation \cite{Creminelli:2006xe,Cheung:2007st} as well as the Ho\v{r}ava gravity \cite{Horava:2009uw,Blas:2009qj} are examples of spatially covariant gravity theories.

One may view the generally covariant scalar-tensor theory and spatially covariant gravity theory as the two faces of the same theory.
When the scalar field possesses a timelike gradient, one is able to fix the so-called unitary gauge with $t=t(\phi)$ (or equivalently $\phi=\phi(t)$) such that the resulting theory takes the form of spatially covariant gravity.
Inversely, the general covariance can be recovered by complementing the spatially covariant gravity with a scalar field, which results in a generally covariant scalar-tensor theory.
These gauge-fixing and gauge-recovering (also dubbed as the St\"{u}ckelberg trick) procedures can be used to transfer from one type of theories to the other.

Although it might be involved to construct ghostfree scalar-tensor theory directly, in the framework of spatially covariant gravity, however, it is relatively straightforward (or even na\"{i}ve) to build the theory with at most three degrees of freedom  \cite{Gao:2014soa,Gao:2014fra,Fujita:2015ymn,Gao:2018znj,Gao:2019lpz,Gao:2018izs,Gao:2019twq}.
We may thus use the spatially covariant gravity as the ``generator'' of ghostfree higher derivative scalar-tensor theory, in particular, with derivatives beyond the second order.
More interestingly, a large class of spatially covariant gravity theories that have $c_{\mathrm{T}} = 1$ has been explored in \cite{Gao:2019liu}, which indicates the possibility that there are more exotic higher derivative scalar-tensor theories that can pass the observational tests.

This work is the first step of attempting fully addressing these issues.
In particular, we focus on the transferring from the spatially covariant gravity to the generally covariant scalar-tensor theory.
When going to the higher order in derivatives, the number and types of the corresponding terms dramatically increase.
The first task is to exhaust all the possible terms that can be included in the Lagrangian and to make a classification of them. 
The generally covariant scalar-tensor monomials are investigated and classified in \cite{Gao:2020juc}.
In this work, we shall make a complimentary classification of the monomials in the spatially covariant gravity.

The paper is organized as following.
In Sec. \ref{sec:corr} we derive the correspondence between the generally covariant derivatives of the scalar field and the spatially covariant geometric quantities.
In Sec. \ref{sec:cls} we make a systematic classification of monomials in the spatially covariant gravity up to $d=4$ in total number of derivatives. This classification is made in order to make the correspondence between two types of theories transparent.
In Sec. \ref{sec:covdec} we derive the covariant 3+1 decomposition of scalar-tensor terms that correspond to the acceleration and extrinsic curvature as well as their spatial derivatives without fixing any gauge.
Sec. \ref{sec:con} concludes.

\smallskip{}

\emph{Notations}: Throughout this work, when writing explicitly, superscripts ``${}^{4}$'' and ``${}^{3}$'' denote spacetime and spatial geometric quantities, respectively.
While $R_{ij}$ and $\nabla_{i}$ always denote the spatial Ricci tensor and spatial covariant derivative.

\section{The correspondence} \label{sec:corr}

In this section, we describe the correspondence between the generally covariant scalar-tensor theory and spatially covariant gravity.

\subsection{Two faces of the scalar-tensor theory}

Generally, the scalar-tensor theory is referred to a theory of which the action takes the form
	\begin{equation}
		S=\int\mathrm{d}^{4}x\sqrt{-g}\,\mathcal{L}\left(g_{ab},\varepsilon_{abcd};\phi,\,{}^{4}\!R_{abcd};\nabla_{a}\right), \label{S_gencov}
	\end{equation}
where the Lagrangian is built of the scalar field $\phi$ and spacetime curvature tensor ${}^{4}\!R_{abcd}$ as well as their covariant derivatives.
The possible appearance of the 4-dimension Levi-Civita tensor $\varepsilon_{abcd}$ signals the parity violation.
In the past few decades, much attention was paid to finding subclasses of (\ref{S_gencov}), in which there are at most 3 degrees of freedom are propagating, i.e., there is at most one additional scalar mode comparing with GR.
The Horndeski theory \cite{Horndeski:1974wa,Nicolis:2008in,Deffayet:2011gz,Kobayashi:2011nu} and the more general degenerate higher order theory \cite{Gleyzes:2014dya,Gleyzes:2014qga,Langlois:2015cwa,Motohashi:2016ftl} are the results. They belong to the subclass of theories of which the action takes the schematic form
	\begin{equation}
		S=\int\mathrm{d}^{4}x\sqrt{-g}\left(\mathcal{A}^{abcd}\,{}^{4}\!R_{abcd}+\mathcal{B}\right),
	\end{equation}
in which the curvature enters linearly, and $\mathcal{A}^{abcd}$ and $\mathcal{B}$ are functions of $\phi$ and its derivatives up to the second order, i.e., of $\{\phi,\nabla_{a}\phi,\nabla_{a}\nabla_{b}\phi\}$ as well as $g_{ab}$ and $\varepsilon_{abcd}$.

One natural question is how to build a theory of which the action takes the schematic form
	\begin{equation}
		S=\int\mathrm{d}^{4}x\sqrt{-g}\left(\mathcal{A}^{a_{1}b_{1}c_{1}d_{1},a_{2}b_{2}c_{2}d_{2}}\,{}^{4}\!R_{a_{1}b_{1}c_{1}d_{1}}\,{}^{4}\!R_{a_{2}b_{2}c_{2}d_{2}}+\mathcal{B}^{abcd}\,{}^{4}\!R_{abcd}+\mathcal{C}\right), \label{S_R2}
	\end{equation}
which contains quadratic curvature terms, and $\mathcal{A}^{a_{1}b_{1}c_{1}d_{1},a_{2}b_{2}c_{2}d_{2}}$, $\mathcal{B}^{abcd}$ and $\mathcal{C}$ are general functions containing derivatives of the scalar field up to the third order, i.e., of $\left\{ \phi,\nabla_{a}\phi,\nabla_{a}\nabla_{b}\phi,\nabla_{a}\nabla_{b}\nabla_{c}\phi\right\}$ as well as $g_{ab}$ and $\varepsilon_{abcd}$.
It is clear that (\ref{S_R2}) includes the quadratic curvature gravity such as the Chern-Simons gravity \cite{Lue:1998mq,Jackiw:2003pm} and the recently studied ghostfree quadratic gravity terms \cite{Deruelle:2012xv,Crisostomi:2017ugk} as special cases.
As has being argued in \cite{Gao:2020juc}, the quadratic curvature terms and derivative terms of the scalar field up to the third order are of the same importance, and thus should be treated in the same footing. 
Moreover, from the lesson of Horndeski theory and degenerate theories, curvature terms and higher derivatives of the scalar field are complimentary to each other.
It is thus very possible that neither is ghostfree but only a combination of both will yield ghostfree theories.

The question is thus how to build such kind of theories.
As being discussed in the Introduction, instead of finely tuning the structure of the Lagrangian such that the higher derivatives are degenerate, there is an alternative approach to the scalar-tensor theories, which is not only technically simpler but also more physically motivated. 
The idea is that as long as the scalar field has a nonvanishing vacuum expectation value, its existence breaks the spacetime general covariance.
The fundamental picture is now a foliation structure of the spacetime, in which the hypersurfaces are specified by the scalar field.
Accordingly, the basic building blocks are thus geometric quantities respecting symmetries of the foliation structure.
To be precise, the hypersurfaces are specified by the scalar field $\phi$, which is now encoded in its gradient $\nabla_{a}\phi  = \epsilon\, N n_{a}$. The normal vector $n_{a}$ is normalized to be $n_{a}n^{a} = \epsilon$, where $\epsilon= -1/+1$ if the gradient of the scalar field is time/spacelike and thus the time/space diffeomorphism is broken.
The action of the theory is thus
	\begin{equation}
		S=\int\mathrm{d}^{4}x\, N\sqrt{h}\, \mathcal{L}\left(\phi,N,h_{ab},\varepsilon_{abcd},{}^{3}\!R_{ab};\mathrm{D}_{a},\pounds_{\bm{n}}\right), \label{S_brane}
	\end{equation}
where $h_{ab} = g_{ab} -\epsilon\, n_{a}n_{b}$ is the induced metric on the hypersurfaces, ${}^{3}\!R_{ab}$ is the intrinsic curvature of the hypersurfaces, $\mathrm{D}_{a}$ is the ``intrinsic'' derivatives that is compatible with $h_{ab}$, the Lie derivative with respect to the normal vector $\pounds_{\bm{n}}$ can be viewed as the ``extrinsic'' derivative.

In this work, we focus on the case where the scalar field is timelike, i.e., we from now on we choose $\epsilon = -1$.
The intrinsic derivative $\mathrm{D}_{a}$ is thus spatial derivative on the hypersurface, and the extrinsic derivative $\pounds_{\bm{n}}$ becomes the temporal derivative.
Non-degenerate higher temporal derivative will cause extra ghostlike degree(s) of freedom. 
Thus we consider subclass of (\ref{S_brane}) with only first order temporal derivative. The first Lie derivative of the induced metric is nothing but the extrinsic curvature $\pounds_{\bm{n}}h_{ab} = 2K_{ab}$. 
The first Lie derivative of the lapse function $\pounds_{\bm{n}}N$ has also been discussed in \cite{Gao:2018znj,Gao:2019lpz}.
It was found that in order to keep the number of degrees of freedom up to 3, there must be constraints on the functional dependence of the Lagrangian on $K_{ab}$ and $\pounds_{\bm{n}}N$. The resulting Lagrangian, at least in some examples, can be transformed to the action containing $K_{ab}$ only by disformal transformation.
In light of this consideration, in this work we focus on the subclass of theories with action of the form
	\begin{equation}
		S=\int\mathrm{d}^{4}x\,N\sqrt{h}\,\mathcal{L}\left(\phi,N,h_{ab},\varepsilon_{abcd},{}^{3}\!R_{ab},K_{ab};\mathrm{D}_{a}\right).
	\end{equation}
We end this subsection by making two comments.
First the shift vector $N_{a}$ that is familiar in the 3+1 formalism is not included explicitly, as 
which itself is not a genuine geometric quantity of the foliation structure and merely characterizes the gauge freedom of choosing a time direction  through $t_a = N n_a + N_a$.  In fact, including terms such as $N_{a} N^{a}$ would introduce more degrees of freedom which may or may not be ghostlike\footnote{This is, however, the idea of building Lorentz breaking massive gravity theories.}.
Second, readers who are familiar with the Einstein-aether theory \cite{Jacobson:2000xp}, which is an effective theory describing a time-like unit vector field coupled to gravity, may soon recognize their similarity to each other.
The main difference is that in our formalism the vector $n_{a}$ is hypersurface orthogonal.

\subsection{From Scalar field to the hypersurface}

In this subsection, we describe the 3+1 decomposition of generally covariant scalar-tensor theory with respect to a general foliation of spacetime. We emphasize that we have not fixed any specific coordinate.
The starting point is a unit timelike vector field $n_{a}$ with $n_{a}n^{a}=-1$, which splits the 4 dimensional spacetime into the tangent and orthogonal parts.
For our purpose, this timelike vector field is also assumed to be hypersurface orthogonal, and thus the induce metric which projects any tensor on the spatial hypersurface is
	\begin{equation}
		h_{ab} \equiv g_{ab}+n_{a}n_{b}.
	\end{equation}

With this setting, all the 4 dimensional covariant object are split into parts that are orthogonal and tangent to the spatial hypersurface. 
The decomposition of the spacetime Riemann tensor yields the Gauss-Codazzi-Ricci equations.
For our purpose, we derive the decomposition of the covariant derivatives of a scalar field, which is not necessarily the scalar field that specifies the hypersurfaces. For the first derivative we have
	\begin{equation}
	\nabla_{a}\phi=-n_{a}\pounds_{\bm{n}}\phi+\mathrm{D}_{a}\phi,\label{nabla_phi_dec}
	\end{equation}
where $\pounds_{\bm{n}}$ stands for the Lie derivative with respect to $n^{a}$, $\mathrm{D}_{a}$ is the projected derivative defined by
	\begin{equation}
		\mathrm{D}_{a}\phi := h_{a}^{a'}\nabla_{a'}\phi,
	\end{equation}
which is also the covariant derivative compatible with $h_{ab}$.
The second order derivative of the scalar field can be decomposed to be\footnote{Throughout this paper, symmetrization is normalized, e.g., $n_{(a}B_{b)}\equiv\frac{1}{2}\left(n_{a}B_{b}+n_{b}B_{a}\right)$ etc.}
	\begin{equation}
	\nabla_{a}\nabla_{b}\phi=n_{a}n_{b}A-2n_{(a}B_{b)}+\Delta_{ab},\label{nabla2_sca_dec_full_sim}
	\end{equation}
with
\begin{eqnarray}
A & \equiv & \pounds_{\bm{n}}^{2}\phi-a^{c}\mathrm{D}_{c}\phi,\label{nabla2_sca_dec_A}\\
B_{a} & \equiv & -a_{a}\pounds_{\bm{n}}\phi+\pounds_{\bm{n}}\mathrm{D}_{a}\phi-K_{a}^{c}\mathrm{D}_{c}\phi,\label{nabla2_sca_dec_B}\\
\Delta_{ab} & \equiv & -K_{ab}\pounds_{\bm{n}}\phi+\mathrm{D}_{a}\mathrm{D}_{b}\phi,\label{nabla2_sca_dec_Delta}
\end{eqnarray}
where $B_{a}$ and $\Delta_{ab}$ are both tangent tensors, i.e., $B_{a}n^a=0$ and $\Delta_{ab}n^b=0$. 
In the above, $a_{a}$ and $K_{ab}$ are the acceleration and the extrinsic curvature defined by 
	\begin{eqnarray}
		a_{a} & \equiv & \pounds_{\bm{n}}n_{a}, \label{a_def_n}\\
		K_{ab} & \equiv & \frac{1}{2}\pounds_{\bm{n}}h_{ab}, \label{K_def_n}
	\end{eqnarray} 
respectively.
For the third order derivative of the scalar field, we find
	\begin{eqnarray}
	\nabla_{c}\nabla_{a}\nabla_{b}\phi & = & -n_{c}n_{a}n_{b}\,U+2n_{c}n_{(a}V_{b)}+n_{a}n_{b}W_{c}\nonumber \\
	&  & -n_{c}X_{ab}-2Y_{c(a}n_{b)}+Z_{cab},\label{nabla3_sca_dec_form}
	\end{eqnarray}
with
\begin{eqnarray}
U & = & \pounds_{\bm{n}}A-2a^{d}B_{d},\label{nabla3_sca_dec_U_form}\\
V_{b} & = & -a_{b}A+\pounds_{\bm{n}}B_{b}-K_{b}^{d}B_{d}-\Delta_{db}a^{d},\label{nabla3_sca_dec_V_form}\\
W_{c} & = & \mathrm{D}_{c}A-2K_{c}^{d}B_{d},\label{nabla3_sca_dec_W_form}\\
X_{ab} & = & -2a_{(a}B_{b)}+\pounds_{\bm{n}}\Delta_{ab}-2K_{(a}^{d}\Delta_{b)d},\label{nabla3_sca_dec_X_form}\\
Y_{cb} & = & -K_{cb}A+\mathrm{D}_{c}B_{b}-K_{c}^{d}\Delta_{db},\label{nabla3_sca_dec_Y_form}\\
Z_{cab} & = & -2K_{c(a}B_{b)}+\mathrm{D}_{c}\Delta_{ab},\label{nabla3_sca_dec_Z_form}
\end{eqnarray}
where $A$, $B_{a}$ and $\Delta_{ab}$ are defined in (\ref{nabla2_sca_dec_A})-(\ref{nabla2_sca_dec_Delta}). The explicit expressions for $U$, $V_{b}$ etc. are given in Appendix \ref{app:xpld3sf} due to their length.

\subsubsection{Unitary gauge}

In the above, the normal vector $n_{a}$ is an arbitrary unit timelike vector field that is hypersurface orthogonal. 
We are free to choose any normal vector $n_{a}$.
When studying the scalar-tensor theory, it is convenient to choose the normal vector to be proportional to the gradient of the scalar field itself, that is,
	\begin{equation}
	n_{a} \rightarrow u_{a} \equiv -\frac{1}{\sqrt{2X}}\nabla_{a}\phi,\label{ua_ug_sf}
	\end{equation}
with
	\begin{equation}
	X:=-\frac{1}{2}\nabla_{a}\phi\nabla^{a}\phi, \label{X_def}
	\end{equation}
which satisfies the normalization $u_{a}u^{a}=-1$.
Decomposition with respect to $u_{a}$ corresponds to the so-called ``unitary gauge''. 
We emphasize that fixing the unitary gauge is merely a special choice of normal vector, which by itself has nothing to do with any coordinate system\footnote{At times, the phrase ``unitary gauge'' indicates the specific coordinates adapted to the hypersurface of the uniform scalar field in the literature.}.

In the unitary gauge, i.e., when being decomposed with respect to hypersurfaces specified by the scalar field itself, the decomposition of the derivatives of the scalar field are dramatically simplified. 
All the spatial derivatives of the scalar field drop out since
	\begin{equation}
	\overset{\mathrm{u}}{\mathrm{D}}_{a}\phi\equiv\overset{\mathrm{u}}{h}_{a}{}^{a'}\nabla_{a}\phi=0,
	\end{equation}
where $\overset{\mathrm{u}}{h}_{ab}$ is defined by
	\begin{equation}
	\overset{\mathrm{u}}{h}_{ab} \equiv  g_{ab}+u_{a}u_{b} .\label{Hab_ug_sf}
	\end{equation}
Here and throughout this paper, a superscript ``${}^{\mathrm{u}}$'' denotes quantities defined with respect to $u_{a}$, which is related to the scalar field through (\ref{ua_ug_sf}).

The first derivative of the scalar field (\ref{nabla_phi_dec}) is thus written as $\nabla_a \phi = -u_{a}/N$ (i.e., (\ref{ua_ug_sf})), where we introduce
	\begin{equation}
	N\equiv\frac{1}{\sqrt{2X}}.\label{lapse_ug_sf}
	\end{equation}
The second derivative of the scalar field (\ref{nabla2_sca_dec_full_sim}) becomes
	\begin{equation}
	\nabla_{a}\nabla_{b}\phi=-\frac{1}{N}\left(u_{a}u_{b}\rho-2u_{(a}\overset{\mathrm{u}}{a}_{b)}+\overset{\mathrm{u}}{K}_{ab}\right),\label{nabla2_Phi_dec_ug}
	\end{equation}
with
	\begin{equation}
		\rho := \pounds_{\bm{u}}\ln N, \label{rho_def}
	\end{equation}
where the acceleration and the extrinsic curvature are now with respect to $u_{a}$, i.e.,
	\begin{eqnarray}
	\overset{\mathrm{u}}{a}_{a} & \equiv  & \pounds_{\bm{u}}u_{a},\\
	\overset{\mathrm{u}}{K}_{ab} & \equiv  & \frac{1}{2}\pounds_{\bm{u}} \overset{\mathrm{u}}{h}_{ab},
	\end{eqnarray}
which are quantities with respect to $u_{a}$ instead of $n_{a}$.
The decomposition of the third derivative (\ref{nabla3_sca_dec_form}) becomes \cite{Gao:2014soa,Gao:2014fra}
	\begin{equation}
	\nabla_{c}\nabla_{a}\nabla_{b}\phi=-u_{c}u_{a}u_{b}\,\overset{\mathrm{u}}{U}+3u_{(c}u_{a}\overset{\mathrm{u}}{V}_{b)}-u_{c}\overset{\mathrm{u}}{X}_{ab}-2\overset{\mathrm{u}}{Y}_{c(a}u_{b)}+\overset{\mathrm{u}}{Z}_{cab},
	\end{equation}
with
	\begin{eqnarray}
	\overset{\mathrm{u}}{U} & = & \frac{1}{N}\left(\rho^{2}-\pounds_{\bm{u}}\rho+2\overset{\mathrm{u}}{a}{}^{d\,}\overset{\mathrm{u}}{a}_{d}\right),\label{nabla3_sca_dec_U_ug}\\
	\overset{\mathrm{u}}{V}_{b} & = & \frac{1}{N}\left(2\rho\,\overset{\mathrm{u}}{a}_{b}-\pounds_{\bm{u}}\overset{\mathrm{u}}{a}_{b}+2\overset{\mathrm{u}}{a}_{d}\,\overset{\mathrm{u}}{K}_{b}{}^{d}\right),\label{nabla3_sca_dec_V_ug}\\
	\overset{\mathrm{u}}{X}_{ab} & = & \frac{1}{N}\left(\rho\,\overset{\mathrm{u}}{K}_{ab}+2\overset{\mathrm{u}}{a}_{a}\overset{\mathrm{u}}{a}_{b}+2\overset{\mathrm{u}}{K}_{a}{}^{d}\overset{\mathrm{u}}{K}_{bd}-\pounds_{\bm{u}}\overset{\mathrm{u}}{K}_{ab}\right),\label{nabla3_sca_dec_X_ug}\\
	\overset{\mathrm{u}}{Y}_{cb} & = & \frac{1}{N}\left[\rho\,\overset{\mathrm{u}}{K}_{cb}-\overset{\mathrm{u}}{\mathrm{D}}_{c}\overset{\mathrm{u}}{a}_{b}+\overset{\mathrm{u}}{a}_{c}\overset{\mathrm{u}}{a}_{b}+\overset{\mathrm{u}}{K}_{c}{}^{d}\overset{\mathrm{u}}{K}_{db}\right],\label{nabla3_sca_dec_Y_ug}\\
	\overset{\mathrm{u}}{Z}_{cab} & = & \frac{1}{N}\left(-\overset{\mathrm{u}}{\mathrm{D}}_{c}\overset{\mathrm{u}}{K}_{ab}+3\overset{\mathrm{u}}{a}_{(c}\overset{\mathrm{u}}{K}_{ab)}\right),\label{nabla3_sca_dec_Z_ug}
	\end{eqnarray}
where $\rho$ is defined in (\ref{rho_def}).

By employing these relations together with the Gauss-Codazzi-Ricci relations, any scalar-tensor term can be recast in terms of geometric quantities of the foliation, as being performed for the Horndeski theory in \cite{Gleyzes:2013ooa}.

\subsection{From the hypersurface to the scalar field} \label{sec:scg2st}

For our purpose, since we shall use the spatially covariant gravity to generate covariant scalar-tensor theories, we also need the inverse procedure, i.e., to reverse the geometric quantities of the foliation in terms of generally covariant scalar-tensor terms. 
This procedure can be traced back to the covariant formulation of Ho\v{r}ava gravity \cite{Germani:2009yt,Blas:2009yd,Jacobson:2010mx,Blas:2010hb} (see also \cite{Chagoya:2018yna}), and is sometimes dubbed as the ``St\"{u}ckelberg trick'' in the literature.

For example, (\ref{ua_ug_sf}) and (\ref{lapse_ug_sf}) can be viewed as the generally covariant correspondence of $u_{a}$ and $N$, respectively.
The generally covariant correspondence of the extrinsic curvature and the acceleration are thus
	\begin{equation}
	\overset{\mathrm{u}}{K}_{ab} = -\frac{1}{\sqrt{2X}}\overset{\mathrm{u}}{h}_{aa'}\,\overset{\mathrm{u}}{h}_{bb'}\nabla^{a'}\nabla^{b'}\phi,\label{Kab_ug_sf}
	\end{equation}
and
	\begin{equation}
	\overset{\mathrm{u}}{a}_{a} = \frac{1}{2X}\overset{\mathrm{u}}{h}_{aa'}\nabla^{b'}\phi\nabla^{a'}\nabla_{b'}\phi.\label{acce_ug_sf_alt}
	\end{equation}
Here $\overset{\mathrm{u}}{h}_{ab}$ is defined in (\ref{Hab_ug_sf}), which now should be understood as
	\begin{equation}
	\overset{\mathrm{u}}{h}_{ab} = g_{ab}+\frac{1}{2X}\nabla_{a}\phi\nabla_{b}\phi.\label{Hab_sf}
	\end{equation}
The spatial Ricci tensor corresponds to
	\begin{equation}
	{}^{3}\!\overset{\mathrm{u}}{R}_{ab} = \overset{\mathrm{u}}{h}_{aa'}\,\overset{\mathrm{u}}{h}_{bb'}\,\overset{\mathrm{u}}{h}_{cd}\,{}^{4}\!R^{a'cb'd}-\frac{1}{2X}\overset{\mathrm{u}}{h}_{aa'}\left(\overset{\mathrm{u}}{h}_{bb'}\,\overset{\mathrm{u}}{h}_{cd}-\overset{\mathrm{u}}{h}_{bc}\,\overset{\mathrm{u}}{h}_{b'd}\right)\nabla^{a'}\nabla^{b'}\phi\nabla^{c}\nabla^{d}\phi.\label{RicciT_3d_ug_sf}
	\end{equation}

For our purpose, we also evaluate 
	\begin{equation}
	\overset{\mathrm{u}}{\mathrm{D}}_{c}\overset{\mathrm{u}}{K}_{ab} = -\frac{1}{\sqrt{2X}}\overset{\mathrm{u}}{h}{}_{c}^{\,c'}\,\overset{\mathrm{u}}{h}{}_{a}^{\,a'}\,\overset{\mathrm{u}}{h}{}_{b}^{\,b'}\left(\nabla_{c'}\nabla_{a'}\nabla_{b'}\phi+\frac{3}{2X}\nabla^{d'}\phi\nabla_{d'}\nabla_{(c'}\phi\nabla_{a'}\nabla_{b')}\phi\right).\label{DcKab_ug_sf}
	\end{equation}
and
	\begin{eqnarray}
	\overset{\mathrm{u}}{\mathrm{D}}_{c}\overset{\mathrm{u}}{a}_{b} & = & \frac{1}{2X}\overset{\mathrm{u}}{h}{}_{c}^{\,c'}\,\overset{\mathrm{u}}{h}{}_{b}^{\,b'}\nabla^{a}\phi\nabla_{c'}\nabla_{b'}\nabla_{a}\phi\nonumber \\
	&  & +\frac{1}{2X}\overset{\mathrm{u}}{h}{}_{c}^{\,c'}\left(\overset{\mathrm{u}}{h}{}_{b}^{\,b'}\,\overset{\mathrm{u}}{h}{}^{ad}+\frac{1}{2X}\overset{\mathrm{u}}{h}{}_{b}^{\,b'}\nabla^{a}\phi\nabla^{d}\phi+\frac{1}{2X}\overset{\mathrm{u}}{h}{}_{b}^{\,a}\nabla^{b'}\phi\nabla^{d}\phi\right)\nabla_{c'}\nabla_{a}\phi\nabla_{b'}\nabla_{d}\phi,\label{Dacce_ug_sf}
	\end{eqnarray}
where again $\overset{\mathrm{u}}{h}_{ab}$ is given in (\ref{Hab_ug_sf}).

Using these relations as well as the Gauss-Codazzi-Ricci equations, the generally covariant scalar-tensor terms can be easily derived from a given spatially covariant term.

\section{Classification of monomials in spatially covariant gravity} \label{sec:cls}

In this section, we make a systematic classification of monomials in the spatially covariant gravity.
These monomials are scalars under spatial diffeomorphism, which are built of the extrinsic and intrinsic curvature $K_{ij}$ and $R_{ij}$, the lapse function $N$, as well as their spatial derivatives.
This classification not only is due to the large number of terms when going to higher order in derivatives, but also makes the correspondence between the spatially covariant gravity and the generally covariant scalar-tensor terms transparent\footnote{There can be different classification with different purpose, see (e.g.) \cite{Zhu:2011yu}.}.

We shall classify various terms and monomials by the derivatives in their corresponding scalar-tensor expressions. 
According to the results in Sec. \ref{sec:scg2st}, schematically we may write
	\begin{equation}
		K_{ij}\sim a_{i}\sim\frac{1}{\nabla\phi}\nabla\nabla\phi,\qquad R_{ij}\sim\frac{1}{\left(\nabla\phi\right)^{2}}\left(\nabla\nabla\phi\right)^{2}\sim{}^{4}\!R, \label{KaR_order}
	\end{equation}
where ${}^{4}\!R$ schematically represents the spacetime curvature tensor, while
	\begin{equation}
		\nabla_{k}K_{ij}\sim\nabla_{j}a_{i}\sim\frac{1}{\nabla\phi}\nabla\nabla\nabla\phi\sim\frac{1}{\left(\nabla\phi\right)^{2}}\left(\nabla\nabla\phi\right)^{2}. \label{DKa_order}
	\end{equation}
In \cite{Gao:2020juc}, scalar-tensor monomials are classified according to the number of derivatives of the scalar field. 
In particular, we may assign each scalar-tensor monomial a set of integers $(c_{0},c_{1},c_{2},\cdots;d_{1},d_{2},d_{3},\cdots)$ in which $c_{m}$ is the number of $m$-th covariant derivatives of the spacetime curvature tensor, $d_{n}$ is the number of $n$-th derivative of the scalar field.
We refer to \cite{Gao:2020juc} for the detailed description.
According to the correspondence in Sec. \ref{sec:scg2st}, we may also assign the same set of integers to the spatially covariant geometric quantities.
For example, $K_{ij}$ and $a_{i}$ correspond to $(c_{0},\cdots;d_{1},d_{2},\cdots) = (0,\cdots;-1,1,\cdots)$, where $d_{1}= -1$ simply stands for a minus power of $\nabla\phi$.
Similarly, ${}^{3}\!R_{ij}$ corresponds to $(c_{0},\cdots;d_{1},d_{2},\cdots) = (0,\cdots;-2,2,\cdots)$ or $(1,\cdots;0,0,\cdots)$.
Moreover, $\nabla_{k}K_{ij}$ and $\nabla_{j}a_{i}$ thus correspond to $(c_{0},\cdots;d_{1},d_{2},d_{3},\cdots) = (0,\cdots;-1,0,1,\cdots)$ or $(0,\cdots;-2,2,0,\cdots)$.
As being argued in \cite{Gao:2020juc}, we treat monomials that are of the same integer $d$ defined by
	\begin{equation}
	d \equiv \sum_{n=0}\left[\left(n+2\right)c_{n}+\left(n+1\right)\,d_{n+2}\right], \label{d_def}
	\end{equation}
as of the same order.
Since $d_{1}$ completely drops out in $d$, from now on we may suppress $d_{1}$ and write $(c_{0},c_{1},\cdots;d_{2},d_{3},\cdots)$.
At this point, it is clear that $d$ is in fact the total number of derivatives in the spatially covariant gravity.

In this work, we will consider monomials up to $d=4$. From (\ref{d_def}) only the first few integers are needed.
Precisely, we will assign each monomial a set of 6 integers $(c_{0},c_{1},c_{2};d_{2},d_{3},d_{4})$.
As the result, we note that (\ref{KaR_order}) and (\ref{DKa_order}) correspond to
	\begin{eqnarray}
	K_{ij}\sim a_{i} & \sim & \left(0,0,0;1,0,0\right),\\
	R_{ij} & \sim & \left(0,0,0;2,0,0\right)\sim\left(1,0,0;0,0,0\right),\\
	\nabla_{k}K_{ij}\sim\nabla_{j}a_{i} & \sim & \left(0,0,0;0,1,0\right).
	\end{eqnarray}
In the rest part of this section, we shall exhaust all the monomials up to $d=4$, and classify these monomials with the set of integers $(c_{0},c_{1},c_{2};d_{2},d_{3},d_{4})$.

\subsection{$d=1$}

The cases of $d=1$ and $d=2$ are simple, of which the monomials are given in Table \ref{tab:mono_d12}.
\begin{table}[H]
	\begin{centering}
		\begin{tabular}{|c|c|l|l|l|c|}
			\hline 
			$d$ & $\#_{\nabla}$ & Form & Irreducible & Reducible & $\left(c_{0},c_{1},c_{2};d_{2},d_{3},d_{4}\right)$\tabularnewline
			\hline 
			\multirow{2}{*}{1} & \multirow{2}{*}{0} & $K$ & $K$ & - & \multirow{2}{*}{$\left(0,0,0;1,0,0\right)$}\tabularnewline
			\cline{3-5} \cline{4-5} \cline{5-5} 
			&  & $a$ & - & - & \tabularnewline
			\hline 
			\multirow{6}{*}{2} & \multirow{4}{*}{0} & $KK$ & $K_{ij}K^{ij},\quad K^{2}$ & - & \multirow{3}{*}{$\left(0,0,0;2,0,0\right)$}\tabularnewline
			\cline{3-5} \cline{4-5} \cline{5-5} 
			&  & $Ka$ & - & - & \tabularnewline
			\cline{3-5} \cline{4-5} \cline{5-5} 
			&  & $aa$ & $a_{i}a^{i}$ & - & \tabularnewline
			\cline{3-6} \cline{4-6} \cline{5-6} \cline{6-6} 
			&  & $R$ & $R$ & - & $\left(1,0,0;0,0,0\right)$\tabularnewline
			\cline{2-6} \cline{3-6} \cline{4-6} \cline{5-6} \cline{6-6} 
			& \multirow{2}{*}{1} & $\nabla K$ & - & - & \multirow{2}{*}{$\left(0,0,0;0,1,0\right)$}\tabularnewline
			\cline{3-5} \cline{4-5} \cline{5-5} 
			&  & $\nabla a$ & - & $\nabla_{i}a^{i}$ & \tabularnewline
			\hline 
		\end{tabular}
		\par\end{centering}
	\caption{Monomials in the spatially covariant gravity with $d=1,2$.}
	\label{tab:mono_d12}
\end{table}

In the case of $d=1$, there is only one term $K\equiv K^{i}_{\phantom{i}i}$, and it is not possible to built a scalar term of $a_{i}$.
Using (\ref{Kab_ug_sf}), after some manipulation, we find that the scalar-tensor correspondence of the monomial $K$ is
	\begin{equation}
	K \rightarrow -\frac{1}{\sqrt{2X}}\left(\square\phi+\frac{1}{2X}\nabla_{a}\phi\nabla_{b}\phi\nabla^{a}\nabla^{b}\phi\right).\label{K_STxpl}
	\end{equation}
Here are throughout this work, we use ``$\rightarrow$'' to denote the scalar-tensor terms that correspond to the monomials (terms) in the spatially covariant gravity.
At the level of Lagrangian, (\ref{K_STxpl}) can be further simplified by integrations by parts.
Using (\ref{ibp_100}), for a general function $f=f(t,N)$, we find (see Appendix \ref{app:ibp_K} for the derivation)
	\begin{equation}
	f\,K \xrightarrow{\sim} \left(F+2X\frac{\partial F}{\partial X}\right)\square\phi-2X\frac{\partial F}{\partial\phi},\label{fK_ST_ibp}
	\end{equation}
	with
	\begin{equation}
	F\left(\phi,X\right) \equiv -\int^{X}\mathrm{d}Y\frac{f\left(\phi,Y\right)}{\left(2Y\right)^{3/2}}, \label{F_def_fK}
	\end{equation}
where $f(\phi,X)$ is understood as the scalar-tensor correspondence of $f(t,N)$.
Throughout this work we use ``$\xrightarrow{\sim}$'' to denote the correspondence from the spatially covariant gravity to the scalar-tensor terms up to total derivatives.
It is thus clear that the trace of the extrinsic curvature $K$ corresponds to nothing but a DGP term \cite{Dvali:2000hr} plus a $k$-essence term, with coefficients being related.

\subsection{$d=2$}

In the case of $d=2$, we find 4 irreducible monomials as shown in the ``Irreducible'' column in Table \ref{tab:mono_d12}. 
The monomial $\nabla_{i}a^{i}$ is reducible in the sense that as $f(t,N)\nabla_{i}a^{i}$ can be reduced to $\tilde{f}(t,N)a_{i}a^{i}$ by integration by parts. 
We focus on the unfactorizable monomials, i.e., those are not product of two or more monomials. 
After some manipulations, we find 
	\begin{eqnarray}
	K_{ij}K^{ij} & \rightarrow & \frac{1}{2X}\nabla_{a}\nabla_{b}\phi\nabla^{a}\nabla^{b}\phi\nonumber \\
	&  & +\frac{1}{2X^{2}}\nabla^{a}\phi\nabla^{b}\phi\nabla_{c}\nabla_{a}\phi\nabla^{c}\nabla_{b}\phi+\frac{1}{8X^{3}}\left(\nabla_{a}\phi\nabla_{b}\phi\nabla^{a}\nabla^{b}\phi\right)^{2}, \label{KijKij_STxpl}
	\end{eqnarray}
	\begin{equation}
	a_{a}a^{a}\rightarrow\frac{1}{4X^{2}}\nabla^{a}\phi\nabla^{b}\phi\nabla_{c}\nabla_{a}\phi\nabla^{c}\nabla_{b}\phi+\frac{1}{8X^{3}}\left(\nabla_{a}\phi\nabla_{b}\phi\nabla^{a}\nabla^{b}\phi\right)^{2}, \label{aa_STxpl}
	\end{equation}
and
	\begin{eqnarray}
	\,{}^{3}\!R & \rightarrow & \,{}^{4}\!R+\frac{1}{X}\,{}^{4}\!R_{ab}\nabla^{a}\phi\nabla^{b}\phi\nonumber \\
	&  & -\frac{1}{2X}\left(\square\phi\right)^{2}-\frac{1}{2X^{2}}\square\phi\nabla_{a}\phi\nabla_{b}\phi\nabla^{a}\nabla^{b}\phi\nonumber \\
	&  & +\frac{1}{2X}\nabla_{a}\nabla_{b}\phi\nabla^{a}\nabla^{b}\phi+\frac{1}{2X^{2}}\nabla^{a}\phi\nabla^{b}\phi\nabla_{c}\nabla_{a}\phi\nabla^{c}\nabla_{b}\phi.\label{R3_ST}
	\end{eqnarray}
The correspondence of $K^2$ (which is factorizable) can be read from (\ref{K_STxpl}) easily.

At this point, it is interesting to show that the combination
	\begin{eqnarray}
	K_{ij}K^{ij}-K^{2} & \rightarrow & -\frac{1}{2X}\Big[\left(\square\phi\right)^{2}-\nabla_{a}\nabla_{b}\phi\nabla^{a}\nabla^{b}\phi\nonumber \\
	&  & \qquad+\frac{1}{X}\square\phi\left(\nabla^{a}\phi\nabla^{b}\phi\nabla_{a}\nabla_{b}\phi\right)-\frac{1}{X}\nabla^{a}\phi\nabla^{b}\phi\nabla_{a}\nabla_{c}\phi\nabla^{c}\nabla_{b}\phi\Big]. \label{KijKij-K2_ST}
	\end{eqnarray}
The right-hand-side is nothing but corresponds to $\mathcal{L}_{(4,0)}$ discussed in \cite{Deffayet:2015qwa} (see eq. (7)).

Similar to the case of $K$, at the level of Lagrangian, (\ref{R3_ST}) can be further reduced by integrations by parts.
For a general function $f(t,N)$, using (\ref{ibp_200}) we get (see Appendix \ref{app:ibp_R3} for the derivation)
	\begin{eqnarray}
	f\,{}^{3}\!R & \xrightarrow{\sim} & f\,{}^{4}\!R+\frac{\partial f}{\partial X}\left(\left(\square\phi\right)^{2}-\nabla_{a}\nabla_{b}\phi\nabla^{a}\nabla^{b}\phi\right)\nonumber \\
	&  & +\left(F+2X\frac{\partial F}{\partial X}\right)\square\phi-2X\frac{\partial F}{\partial\phi}\nonumber \\
	&  & +\frac{1}{2X}\left(f-2X\frac{\partial f}{\partial X}\right)\Big[\left(\square\phi\right)^{2}-\nabla_{a}\nabla_{b}\phi\nabla^{a}\nabla^{b}\phi\nonumber \\
	&  & \qquad\qquad+\frac{1}{X}\nabla^{a}\phi\nabla^{b}\phi\square\phi\nabla_{a}\nabla_{b}\phi-\frac{1}{X}\nabla^{a}\phi\nabla^{b}\phi\nabla_{a}\nabla_{c}\phi\nabla^{c}\nabla_{b}\phi\Big], \label{R3_STibp}
	\end{eqnarray}
with
	\begin{equation}
	F\left(\phi,X\right)\equiv-\int^{X}\mathrm{d}Y\frac{1}{Y}\frac{\partial f\left(\phi,Y\right)}{\partial\phi}. \label{F_def_R3}
	\end{equation}
Note the last two lines take exactly the same form as (\ref{KijKij-K2_ST}). 
It thus immediately follows that the combination
	\begin{eqnarray}
	\frac{\partial\left(Nf\right)}{\partial N}\left(K_{ij}K^{ij}-K^{2}\right)+f\,{}^{3}\!R & \xrightarrow{\sim} & f\,{}^{4}\!R+\frac{\partial f}{\partial X}\left(\left(\square\phi\right)^{2}-\nabla_{a}\nabla_{b}\phi\nabla^{a}\nabla^{b}\phi\right)\nonumber \\
	&  & +\left(F+2X\frac{\partial F}{\partial X}\right)\square\phi-2X\frac{\partial F}{\partial\phi}, \label{KKR3com_ST}
	\end{eqnarray}
where $F$ is given in (\ref{F_def_R3}).
It is interesting to note that the right-hand-side of (\ref{KKR3com_ST}) takes exactly the form of Horndeski Lagrangian.
Setting $f\rightarrow 1$ in (\ref{KKR3com_ST}) yields
	\begin{equation}
	K_{ij}K^{ij}-K^{2}+\,{}^{3}\!R\xrightarrow{\sim}\, {}^{4}\! R,
	\end{equation}
where the left-hand-side recovers the familiar ADM form of the general relativity.

\subsection{$d=3$}

For $d=3$, the monomials are summarized in Table \ref{tab:mono_d3}.
\begin{table}[H]
	\begin{centering}
		\begin{tabular}{|c|c|l|l|l|c|}
			\hline 
			$d$ & $\#_{\nabla}$ & Form & Irreducible & Reducible & $\left(c_{0},c_{1},c_{2};d_{2},d_{3},d_{4}\right)$\tabularnewline
			\hline 
			\multirow{13}{*}{3} & \multirow{6}{*}{0} & $KKK$ & $K_{ij}K^{jk}K_{k}^{i},\quad K_{ij}K^{ij}K,\quad K^{3}$ & - & \multirow{4}{*}{$\left(0,0,0;3,0,0\right)$}\tabularnewline
			\cline{3-5} \cline{4-5} \cline{5-5} 
			&  & $KKa$ & - & - & \tabularnewline
			\cline{3-5} \cline{4-5} \cline{5-5} 
			&  & $Kaa$ & $K_{ij}a^{i}a^{j},\quad Ka_{i}a^{i}$ & - & \tabularnewline
			\cline{3-5} \cline{4-5} \cline{5-5} 
			&  & $aaa$ & - & - & \tabularnewline
			\cline{3-6} \cline{4-6} \cline{5-6} \cline{6-6} 
			&  & $RK$ & $R^{ij}K_{ij},\quad RK$ & - & \multirow{2}{*}{$\left(1,0,0;1,0,0\right)$}\tabularnewline
			\cline{3-5} \cline{4-5} \cline{5-5} 
			&  & $Ra$ & - & - & \tabularnewline
			\cline{2-6} \cline{3-6} \cline{4-6} \cline{5-6} \cline{6-6} 
			& \multirow{5}{*}{1} & $K\nabla K$ & $\varepsilon_{ijk}K_{l}^{i}\nabla^{j}K^{kl}$ & - & \multirow{4}{*}{$\left(0,0,0;1,1,0\right)$}\tabularnewline
			\cline{3-5} \cline{4-5} \cline{5-5} 
			&  & $K\nabla a$ & $K_{ij}\nabla^{i}a^{j},\quad K\nabla_{i}a^{i}$ & - & \tabularnewline
			\cline{3-5} \cline{4-5} \cline{5-5} 
			&  & $a\nabla K$ & - & $a^{j}\nabla_{i}K_{j}^{i},\quad a^{i}\nabla_{i}K$ & \tabularnewline
			\cline{3-5} \cline{4-5} \cline{5-5} 
			&  & $a\nabla a$ & - & - & \tabularnewline
			\cline{3-6} \cline{4-6} \cline{5-6} \cline{6-6} 
			&  & $\nabla R$ & - & - & $\left(0,1,0;0,0,0\right)$\tabularnewline
			\cline{2-6} \cline{3-6} \cline{4-6} \cline{5-6} \cline{6-6} 
			& \multirow{2}{*}{2} & $\nabla\nabla K$ & - & $\nabla^{i}\nabla^{j}K_{ij},\quad\nabla^{2}K$ & \multirow{2}{*}{$\left(0,0,0;0,0,1\right)$}\tabularnewline
			\cline{3-5} \cline{4-5} \cline{5-5} 
			&  & $\nabla\nabla a$ & - & - & \tabularnewline
			\hline 
		\end{tabular}
		\par\end{centering}
	\caption{Monomials of spatially covariant gravity with $d=3$.}
	
	\label{tab:mono_d3}
\end{table}

Although it is relatively simple for $d=1,2$, the scalar-tensor correspondences rapidly become complicated and unreadable when $d$ goes large.
In \cite{Gao:2020juc}, the monomials built of the scalar field and the curvature tensor as well as their covariant derivatives are systematically classified and described.
In particular, up to $d=4$, a notation in the type of $\bm{E}^{(c_{0};d_{2},d_{3})}_{n}$'s are developed to denote such monomials.
We present the concrete expressions of the relevant terms in Appendix \ref{app:STmono} and refer to \cite{Gao:2020juc} for more details. 
For example, with such notations (\ref{K_STxpl}) can be written as
	\begin{equation}
	K \rightarrow -\bm{E}_{1}^{(0;1,0)}-\bm{E}_{2}^{(0;1,0)},\label{K_STm}
	\end{equation}
where $\bm{E}_{1}^{(0;1,0)}$ and $\bm{E}_{2}^{(0;1,0)}$ are defined in (\ref{E010_1}) and (\ref{E010_2}), respectively.
Correspondingly, (\ref{fK_ST_ibp}) can be written as
	\begin{equation}
	f\, K\xrightarrow{\sim}\left(F+\sigma\frac{\partial F}{\partial\sigma}\right)\sigma\bm{E}_{1}^{\left(1,0,0\right)}-\sigma^{2}\frac{\partial F}{\partial\phi},
	\end{equation}
where $F$ is given in (\ref{F_def_fK}) and we used $2X\frac{\partial }{\partial X} \equiv \sigma\frac{\partial }{\partial\sigma}$.
For $d=2$, (\ref{KijKij_STxpl}), (\ref{KijKij-K2_ST}) and (\ref{aa_STxpl}) are written as
	\begin{equation}
	K_{ij}K^{ij}\rightarrow\bm{E}_{1}^{(2,0,0)}+2\bm{E}_{2}^{(2,0,0)}+\left(\bm{E}_{2}^{(1,0,0)}\right)^{2}, \label{KabKab_STm}
	\end{equation}
	\begin{equation}
	K^{2}-K_{ij}K^{ij}\rightarrow-\bm{E}_{1}^{(2,0,0)}-2\bm{E}_{2}^{(2,0,0)}+\left(\bm{E}_{1}^{(1,0,0)}\right)^{2}+2\bm{E}_{1}^{(1,0,0)}\bm{E}_{2}^{(1,0,0)},
	\end{equation}
and
	\begin{equation}
	a_{i}a^{i} \rightarrow \bm{E}_{2}^{(0;2,0)} + \left(\bm{E}_{2}^{(0;1,0)}\right)^{2},\label{aa_STm}
	\end{equation}
respectively.
The combination (\ref{KKR3com_ST}) becomes 
	\begin{eqnarray}
	\frac{\partial\left(Nf\right)}{\partial N}\left(K_{ij}K^{ij}-K^{2}\right)+f\,{}^{3}\!R & \xrightarrow{\sim} & f\,{}^{4}\!R+\sigma\frac{\partial f}{\partial\sigma}\left(\left(\bm{E}_{1}^{\left(1,0,0\right)}\right)^{2}-\bm{E}_{1}^{\left(2,0,0\right)}\right)\nonumber \\
	&  & +\left(F+\sigma\frac{\partial F}{\partial\sigma}\right)\sigma\bm{E}_{1}^{\left(1,0,0\right)}-\sigma^{2}\frac{\partial F}{\partial\phi},
	\end{eqnarray}
with $F$  given in (\ref{F_def_R3}).

In the following, we derive the correspondences for the unfactorizable monomials with $d=3$, as the factorizable terms can be read easily (see Appendix \ref{app:d4fac}).
We find 
	\begin{equation}
	K_{j}^{i}K_{k}^{j}K_{i}^{k}\rightarrow-\bm{E}_{1}^{(0;3,0)}-3\bm{E}_{2}^{(0;3,0)}-3\bm{E}_{2}^{(0;1,0)}\bm{E}_{2}^{(0;2,0)}-\left(\bm{E}_{2}^{(0;1,0)}\right)^{3},\label{KabKbcKca_STm}
	\end{equation}
	\begin{equation}
	K_{ij}a^{i}a^{j}\rightarrow-\bm{E}_{2}^{(0;3,0)}-2\bm{E}_{2}^{(0;1,0)}\bm{E}_{2}^{(0;2,0)}-\left(\bm{E}_{2}^{(0;1,0)}\right)^{3},\label{Kabaaab_STm}
	\end{equation}
	\begin{eqnarray}
	^{3}\!R^{ij}K_{ij} & \rightarrow & -\bm{E}_{1}^{(1;1,0)}-\bm{E}_{2}^{(1;1,0)}-2\bm{E}_{3}^{(1;1,0)}-\bm{E}_{2}^{(1;0,0)}\bm{E}_{2}^{(0;1,0)}\nonumber \\
	&  & -\bm{E}_{1}^{(0;3,0)}-3\bm{E}_{2}^{(0;3,0)}+\bm{E}_{2}^{(0;1,0)}\left(\bm{E}_{1}^{(0;2,0)}-\bm{E}_{2}^{(0;2,0)}\right)\nonumber \\
	&  & +\bm{E}_{1}^{(0;1,0)}\left(\left(\bm{E}_{2}^{(0;1,0)}\right)^{2}+\bm{E}_{1}^{(0;2,0)}+2\bm{E}_{2}^{(0;2,0)}\right),\label{R3abKab_STm}
	\end{eqnarray}
and
	\begin{eqnarray}
	K_{ij}\nabla^{i}a^{j} & \rightarrow & -\bm{E}_{1}^{(0;3,0)}-4\bm{E}_{2}^{(0;3,0)}-\bm{E}_{4}^{(0;1,1)}-2\bm{E}_{5}^{(0;1,1)}\nonumber \\
	&  & -\bm{E}_{2}^{(0;1,0)}\left(\bm{E}_{1}^{(0;2,0)}+7\bm{E}_{2}^{(0;2,0)}+\bm{E}_{3}^{(0;0,1)}+3\left(\bm{E}_{2}^{(0;1,0)}\right)^{2}\right).\label{KabDaab_STm}
	\end{eqnarray}
Note the scalar-tensor correspondence of $K_{ij}\nabla^{i}a^{j}$ (and also of $K \nabla_{i}a^{i}$) contains third order derivative of the scalar field.
There is also a single parity-violating monomial, 
	\begin{equation}
	\varepsilon_{ijk}K_{l}^{i}\nabla^{j}K^{kl}\rightarrow-\bm{F}_{1}^{(0;1,1)} \equiv \frac{1}{2}\bm{F}_{1}^{(1;1,0)}.\label{epsKDK_STm}
	\end{equation}
Note the parity-violating terms $\bm{F}_{1}^{(1;1,0)}$ has been discussed in \cite{Crisostomi:2017ugk} (see eq. (3.12)).

\subsection{$d=4$}

The monomials with $d=4$ are exhausted in Table \ref{tab:mono_d4}.
\begin{table}[H]
	\begin{centering}
		\begin{tabular}{|c|c|l|>{\raggedright}p{6cm}|>{\raggedright}p{5cm}|c|}
			\hline 
			$d$ & $\#_{\nabla}$ & Form & Irreducible & Reducible & $\left(c_{0},c_{1},c_{2};d_{2},d_{3},d_{4}\right)$\tabularnewline
			\hline 
			\multirow{29}{*}{4} & \multirow{9}{*}{0} & $KKKK$ & $K_{ij}K^{jk}K_{k}^{i}K,\quad\left(K_{ij}K^{ij}\right)^{2},$
			
			$K_{ij}K^{ij}K^{2},\quad K^{4}$ & - & \multirow{5}{*}{$\left(0,0,0;4,0,0\right)$}\tabularnewline
			\cline{3-5} \cline{4-5} \cline{5-5} 
			&  & $KKKa$ & - & - & \tabularnewline
			\cline{3-5} \cline{4-5} \cline{5-5} 
			&  & $KKaa$ & $K_{ik}K_{j}^{k}a^{i}a^{j},$
			
			$K_{ij}K^{ij}a_{k}a^{k},\quad K^{2}a_{i}a^{i},\quad KK_{ij}a^{i}a^{j}$ & - & \tabularnewline
			\cline{3-5} \cline{4-5} \cline{5-5} 
			&  & $Kaaa$ & - & - & \tabularnewline
			\cline{3-5} \cline{4-5} \cline{5-5} 
			&  & $aaaa$ & $\left(a_{i}a^{i}\right)^{2}$ & - & \tabularnewline
			\cline{3-6} \cline{4-6} \cline{5-6} \cline{6-6} 
			&  & $RKK$ & $R_{ij}K_{k}^{i}K^{jk},$
			
			$RK_{ij}K^{ij},\quad R_{ij}K^{ij}K,\quad RK^{2}$ & - & \multirow{3}{*}{$\left(1,0,0;2,0,0\right)$}\tabularnewline
			\cline{3-5} \cline{4-5} \cline{5-5} 
			&  & $RKa$ & $\varepsilon_{ijk}R_{l}^{i}K^{jl}a^{k}$ & - & \tabularnewline
			\cline{3-5} \cline{4-5} \cline{5-5} 
			&  & $Raa$ & $R_{ij}a^{i}a^{j},\quad Ra_{i}a^{i}$ & - & \tabularnewline
			\cline{3-6} \cline{4-6} \cline{5-6} \cline{6-6} 
			&  & $RR$ & $R_{ij}R^{ij},\quad R^{2}$ & - & $\left(2,0,0;0,0,0\right)$\tabularnewline
			\cline{2-6} \cline{3-6} \cline{4-6} \cline{5-6} \cline{6-6} 
			& \multirow{10}{*}{1} & $KK\nabla K$ & $\varepsilon_{ijk}\nabla_{m}K_{n}^{i}K^{jm}K^{kn},\quad\varepsilon_{ijk}\nabla^{i}K_{m}^{j}K_{n}^{k}K^{mn},$
			
			$\varepsilon_{ijk}\nabla^{i}K_{l}^{j}K^{kl}K$ & - & \multirow{6}{*}{$\left(0,0,0;2,1,0\right)$}\tabularnewline
			\cline{3-5} \cline{4-5} \cline{5-5} 
			&  & $KK\nabla a$ & $K_{i}^{k}K_{jk}\nabla^{i}a^{j},$
			
			$K_{ij}K^{ij}\nabla_{k}a^{k},\quad KK_{ij}\nabla^{i}a^{j},\quad K^{2}\nabla_{i}a^{i}$ & - & \tabularnewline
			\cline{3-5} \cline{4-5} \cline{5-5} 
			&  & $Ka\nabla K$ & $K_{j}^{i}a^{j}\nabla_{k}K_{i}^{k},\quad K_{j}^{i}a^{j}\nabla_{i}K,$ & $K^{ik}a^{j}\nabla_{j}K_{ik},\quad K^{ik}a^{j}\nabla_{k}K_{ij},$
			
			$Ka^{i}\nabla_{i}K,\quad Ka^{i}\nabla_{j}K_{i}^{j}$ & \tabularnewline
			\cline{3-5} \cline{4-5} \cline{5-5} 
			&  & $Ka\nabla a$ & $\varepsilon_{ijk}K_{l}^{i}a^{j}\nabla^{k}a^{l}$ & - & \tabularnewline
			\cline{3-5} \cline{4-5} \cline{5-5} 
			&  & $aa\nabla K$ & - & $\varepsilon_{ijk}a^{l}a^{i}\nabla^{j}K_{l}^{k}$ & \tabularnewline
			\cline{3-5} \cline{4-5} \cline{5-5} 
			&  & $aa\nabla a$ & $a_{i}a^{i}\nabla_{j}a^{j}$ & $a^{i}a^{j}\nabla_{i}a_{j}$ & \tabularnewline
			\cline{3-6} \cline{4-6} \cline{5-6} \cline{6-6} 
			&  & $R\nabla K$ & $\varepsilon_{ijk}R^{il}\nabla^{j}K_{l}^{k}$ & - & \multirow{2}{*}{$\left(1,0,0;0,1,0\right)$}\tabularnewline
			\cline{3-5} \cline{4-5} \cline{5-5} 
			&  & $R\nabla a$ & $R\nabla_{i}a^{i}$ & $R^{ij}\nabla_{i}a_{j}$ & \tabularnewline
			\cline{3-6} \cline{4-6} \cline{5-6} \cline{6-6} 
			&  & $K\nabla R$ & - & $\varepsilon_{ijk}K^{il}\nabla^{j}R_{l}^{k}$ & \multirow{2}{*}{$\left(0,1,0;1,0,0\right)$}\tabularnewline
			\cline{3-5} \cline{4-5} \cline{5-5} 
			&  & $a\nabla R$ & - & $a^{i}\nabla_{i}R$ & \tabularnewline
			\cline{2-6} \cline{3-6} \cline{4-6} \cline{5-6} \cline{6-6} 
			& \multirow{8}{*}{2} & $\nabla K\nabla K$ & $\nabla_{k}K_{ij}\nabla^{k}K^{ij},\quad\nabla_{i}K^{ij}\nabla_{k}K_{j}^{k},$
			
			$\nabla_{i}K^{ij}\nabla_{j}K,\quad\nabla_{i}K\nabla^{i}K$ & $\nabla_{i}K_{jk}\nabla^{k}K^{ij}$ & \multirow{3}{*}{$\left(0,0,0;0,2,0\right)$}\tabularnewline
			\cline{3-5} \cline{4-5} \cline{5-5} 
			&  & $\nabla K\nabla a$ & - & $\varepsilon_{ijk}\nabla^{i}K_{l}^{j}\nabla^{l}a^{k}$ & \tabularnewline
			\cline{3-5} \cline{4-5} \cline{5-5} 
			&  & $\nabla a\nabla a$ & $\nabla_{i}a_{j}\nabla^{i}a^{j},\quad\left(\nabla_{i}a^{i}\right)^{2}$ & - & \tabularnewline
			\cline{3-6} \cline{4-6} \cline{5-6} \cline{6-6} 
			&  & $K\nabla\nabla K$ & - & $K^{ij}\nabla_{i}\nabla_{j}K,\quad K^{ij}\nabla_{j}\nabla_{k}K_{i}^{k},$
			
			$K^{ij}\nabla_{k}\nabla_{j}K_{i}^{k},\quad K\nabla_{i}\nabla_{j}K^{ij},$
			
			$K^{ij}\nabla^{2}K_{ij},\quad K\nabla^{2}K$ & \multirow{4}{*}{$\left(0,0,0;1,0,1\right)$}\tabularnewline
			\cline{3-5} \cline{4-5} \cline{5-5} 
			&  & $K\nabla\nabla a$ & - & $\varepsilon_{ijk}K_{l}^{i}\nabla^{j}\nabla^{k}a^{l}$ & \tabularnewline
			\cline{3-5} \cline{4-5} \cline{5-5} 
			&  & $a\nabla\nabla K$ & - & $\varepsilon_{ijk}a^{i}\nabla_{l}\nabla^{j}K^{lk},\quad\varepsilon_{ijk}a^{i}\nabla^{j}\nabla_{l}K^{lk},$
			
			$\varepsilon_{ijk}a^{l}\nabla^{i}\nabla^{j}K_{l}^{k}$ & \tabularnewline
			\cline{3-5} \cline{4-5} \cline{5-5} 
			&  & $a\nabla\nabla a$ & - & $a^{i}\nabla_{i}\nabla_{j}a^{j},\quad a^{i}\nabla^{2}a_{i}$ & \tabularnewline
			\cline{3-6} \cline{4-6} \cline{5-6} \cline{6-6} 
			&  & $\nabla\nabla R$ & - & $\nabla^{2}R$ & $\left(0,0,1;0,0,0\right)$\tabularnewline
			\cline{2-6} \cline{3-6} \cline{4-6} \cline{5-6} \cline{6-6} 
			& \multirow{2}{*}{3} & $\nabla\nabla\nabla K$ & - & $\varepsilon_{ijk}\nabla_{l}\nabla^{i}\nabla^{j}K^{kl}$ & 5th der.\tabularnewline
			\cline{3-6} \cline{4-6} \cline{5-6} \cline{6-6} 
			&  & $\nabla\nabla\nabla a$ & - & $\nabla_{i}\nabla^{2}a^{i},\quad\nabla^{2}\nabla_{i}a^{i}$ & 5th der.\tabularnewline
			\hline 
		\end{tabular}
		\par\end{centering}
	\caption{Monomials of spatially covariant gravity with $d=4$.}
	\label{tab:mono_d4}
\end{table}
The scalar-tensor correspondences of monomials with $d=4$ will be presented in a following up paper \cite{Gao:inprogress}.
According to Table \ref{tab:mono_d4}, the third order derivatives of the scalar field also naturally arise.

\section{Covariant 3+1 decomposition} \label{sec:covdec}

In the above we have present the correspondences of the spatially covariant gravity in terms of covariant scalar-tensor terms.
Monomials of spatially covariant gravity correspond to the 3+1 decomposition of scalar-tensor terms after fixing the so-called unitary gauge (more precisely, after choosing the adapted coordinates).
In practise, one may also need the 3+1 decomposition of the covariant expressions without fixing any specific gauge. This is in particular the case if one would like to perform a covariant Hamiltonian analysis.
The purpose of this section is to derive the relevant decomposition in a covariant manner. 
In fact, fixing the so-called unitary gauge is a merely choice of foliation, which is not necessarily related to choosing any specific coordinates, and thus can be done ``covariantly''.

To this end, note the unit timelike vector that is associated with the scalar field is $u_{a}$ defined in (\ref{ua_ug_sf}).
We are free to decompose $u_a$  with respect to an arbitrary foliation with normal vector $n_a$ as
	\begin{equation}
	u_{a}=-n_{a}\alpha+\beta_{a},\qquad\text{with}\quad n^{a}\beta_{a}\equiv0.\label{ua_na_dec}
	\end{equation}
For our purpose we assume $n_{a}$ is timelike such that $n_{a}n^{a} = -1$.
We thus have 
	\begin{equation}
		\beta_{a} = -\frac{\mathrm{D}_{a}\phi}{\sqrt{2X}},
	\end{equation}
with $X$ is defined in (\ref{X_def}), which is decomposed to be
	\begin{equation}
	X=\frac{1}{2}\left(\pounds_{\bm{n}}\phi\right)^{2}-\frac{1}{2}\mathrm{D}_{a}\phi\mathrm{D}^{a}\phi.
	\end{equation}
Note $\mathrm{D}_{a}$ is the projected derivative with respect to $h_{ab}$.
$\alpha$ is related to $\beta_{a}$ through
	\begin{equation}
	\alpha =-\sqrt{1+\beta_{a}\beta^{a}} \equiv -\frac{\pounds_{\bm{n}}\phi}{\sqrt{2X}}.
	\end{equation}
Note here we have implicitly assumed $u_{a}$ to be timelike and thus $X$ defined as in (\ref{X_def}). 
When the scalar field possesses a spacelike gradient, $u_{a}$ becomes spacelike and we simply replace $X\rightarrow -X$.

It is thus clear that the unitary gauge corresponds a special choice of foliation such that $\mathrm{D}_{a}\phi = 0$, i.e., 
	\begin{equation}
	\beta_{a} = 0, \qquad \alpha  = -1.
	\end{equation}
which corresponds simply to choosing $n_{a}\rightarrow u_{a}$.
In this sense, fixing the unitary gauge is to perform the 3+1 decomposition with respect to a special foliation, in which the hypersurfaces are specified by the scalar field itself.

The induced metric $\overset{\mathrm{u}}{h}_{ab}$ with respect to $u_{a}$ is defined as in (\ref{Hab_ug_sf}), which should be understood as the scalar-tensor expression (\ref{Hab_sf}) and can be decomposed to be
	\begin{equation}
	\overset{\mathrm{u}}{h}_{ab} =h_{ab}+n_{a}n_{b}\beta_{c}\beta^{c}-2n_{(a}\beta_{b)}\alpha+\beta_{a}\beta_{b}.\label{ind_Hab_hab}
	\end{equation}
It is thus clear that fixing the unitary gauge implies\footnote{It can be thought as the ``hat'' drops out.}
	\begin{equation}
	\overset{\mathrm{u}}{h}_{ab} \xrightarrow{\; \beta_{a}=0 \;} h_{ab}.
	\end{equation}

The covariant decomposition of the extrinsic curvature $\overset{\mathrm{u}}{K}_{ab}$, which  represents the scalar-tensor expression in (\ref{Kab_ug_sf}), is given by 
	\begin{equation}
	\overset{\mathrm{u}}{K}_{ab}=n_{a}n_{b}\overset{\mathrm{u}}{K}{}^{\perp\perp}-2n_{(a}\overset{\mathrm{u}}{K}{}_{b)}^{\perp\parallel}+\overset{\mathrm{u}}{K}{}_{ab}^{\parallel\parallel},\label{Kab_Hind_dec}
	\end{equation}
with
	\begin{align}
	\overset{\mathrm{u}}{K}{}^{\perp\perp} & =-\frac{1}{\alpha}K^{cd}\beta_{c}\beta_{d}-\frac{1}{\alpha}\beta_{d}\beta^{d}\beta^{c}\pounds_{\bm{n}}\beta_{c}\nonumber \\
	& +a^{d}\beta_{d}\beta_{c}\beta^{c}+\beta^{c}\beta^{d}\mathrm{D}_{c}\beta_{d},\label{Kab_Hind_nn}
	\end{align}
	\begin{eqnarray}
	\overset{\mathrm{u}}{K}{}_{a}^{\perp\parallel} & = & -\frac{1}{2}\beta_{c}\beta^{c}\pounds_{\bm{n}}\beta_{a}-\frac{1}{2}\beta_{a}\beta^{c}\pounds_{\bm{n}}\beta_{c}-K_{a}^{c}\beta_{c}\nonumber \\
	&  & +\frac{\alpha}{2}a_{a}\beta_{c}\beta^{c}+\frac{1}{4\alpha}\mathrm{D}_{a}\left(\beta_{c}\beta^{c}\right)+\frac{\alpha}{2}\beta^{c}\mathrm{D}_{c}\beta_{a}\nonumber \\
	&  & +\frac{\alpha}{2}\beta_{a}a^{c}\beta_{c}+\frac{1}{2\alpha}\beta_{a}\beta^{c}\beta^{d}\mathrm{D}_{c}\beta_{d},\label{Kab_Hind_nh}
	\end{eqnarray}
and
	\begin{eqnarray}
	\overset{\mathrm{u}}{K}{}_{ab}^{\parallel\parallel} & = & -\alpha K_{ab}-\alpha\,\beta_{(a}\pounds_{\bm{n}}\beta_{b)}\nonumber \\
	&  & +\alpha^{2}a_{(a}\beta_{b)}+\mathrm{D}_{(a}\beta_{b)}+\beta^{c}\beta_{(a}\mathrm{D}_{c}\beta_{b)}.\label{Kab_Hind_hh}
	\end{eqnarray}
We emphasize that through out this paper $a_{a}$ and $K_{ab}$ are the acceleration and extrinsic curvature associated with $n_{a}$ defined in (\ref{a_def_n}) and (\ref{K_def_n}).
Similarly, the covariant decomposition of the acceleration $\overset{\mathrm{u}}{a}_{a}$, which stands for the scalar-tensor expression (\ref{acce_ug_sf_alt}), is thus
	\begin{equation}
	\overset{\mathrm{u}}{a}_{a}=-n_{a}\overset{\mathrm{u}}{a}{}^{\perp}+\overset{\mathrm{u}}{a}{}_{a}^{\parallel},\label{acce_Hind_dec}
	\end{equation}
where
	\begin{equation}
	\overset{\mathrm{u}}{a}{}^{\perp} \equiv -\beta^{b}\pounds_{\bm{n}}\beta_{b}+\alpha\,a^{b}\beta_{b}+\frac{1}{\alpha}\beta^{a}\beta^{b}\mathrm{D}_{a}\beta_{b},\label{acce_Hind_perp}
	\end{equation}
and
	\begin{equation}
	\overset{\mathrm{u}}{a}{}_{a}^{\parallel} \equiv \alpha^{2}a_{a}-\alpha\,\pounds_{\bm{n}}\beta_{a}+\beta^{c}\mathrm{D}_{c}\beta_{a}.\label{acce_Hind_parallel}
	\end{equation}

The deviation from the unitary gauge is encoded in the nonvanishing $\beta_{a}$.
We have
	\begin{eqnarray}
		\overset{\mathrm{u}}{K}{}^{\perp\perp} \sim \mathcal{O}(|\beta|^2), \qquad \overset{\mathrm{u}}{K}{}_{a}^{\perp\parallel} \sim \mathcal{O}(|\beta|^1), \qquad 
		\overset{\mathrm{u}}{K}{}_{ab}^{\parallel\parallel} \sim \mathcal{O}(|\beta|^0),
	\end{eqnarray}
and similarly
	\begin{equation}
		\overset{\mathrm{u}}{a}{}^{\perp} \sim \mathcal{O}(|\beta|^1), \qquad \overset{\mathrm{u}}{a}{}_{a}^{\parallel} \sim \mathcal{O}(|\beta|^0).
	\end{equation}
It is clear that fixing the unitary gauge implies
	\begin{eqnarray}
	\overset{\mathrm{u}}{K}_{ab} & \xrightarrow{\;\beta_{a}= 0 \;} & K_{ab}, \\
	\overset{\mathrm{u}}{a}_{a} & \xrightarrow{\;\beta_{a}=0 \;} & a_{a}.
	\end{eqnarray}
On the other hand, generally $\overset{\mathrm{u}}{K}_{ab}$ and $\overset{\mathrm{u}}{a}_{a}$ have timelike components (i.e., components proportional to $n_{a}$). In particular, there are terms involving the Lie derivative $\pounds_{\bm{n}}\beta_{a}$. However, all terms involving the Lie derivatives are proportional to $\beta_{a} \propto \mathrm{D}_{a}\phi $, which are thus vanishing in the unitary gauge.
When deviating from the unitary gauge, such terms may apparently signals additional dynamical degrees of freedom, which are argued to be removable by appropriate spatial boundary conditions \cite{DeFelice:2018ewo}.

For our purpose and late convenience, we also derive the covariant decomposition of $\overset{\mathrm{u}}{\mathrm{D}}_{c}\overset{\mathrm{u}}{K}_{ab}$ and $\overset{\mathrm{u}}{\mathrm{D}}_{a}\overset{\mathrm{u}}{a}_{b}$, which are given by 
	\begin{equation}
	\overset{\mathrm{u}}{\mathrm{D}}_{c}\overset{\mathrm{u}}{K}_{ab} \equiv \overset{\mathrm{u}}{h}{}_{c}^{\,c'}\,\overset{\mathrm{u}}{h}{}_{a}^{\,a'}\,\overset{\mathrm{u}}{h}{}_{b}^{\,b'}\nabla_{c'}\overset{\mathrm{u}}{K}_{a'b'}, \label{cov_corr_DcKab}
	\end{equation}
and
	\begin{equation}
	\overset{\mathrm{u}}{\mathrm{D}}_{a}\overset{\mathrm{u}}{a}_{b} \equiv \overset{\mathrm{u}}{h}{}_{a}^{\,a'}\overset{\mathrm{u}}{h}{}_{b}^{\,b'}\nabla_{a'}\overset{\mathrm{u}}{a}_{b'}, \label{cov_corr_Daab}
	\end{equation}
of which the explicit scalar-tensor expressions are given in (\ref{DcKab_ug_sf}) and (\ref{Dacce_ug_sf}), respectively.	
After some manipulations, we find
	\begin{eqnarray}
	\nabla_{c}\overset{\mathrm{u}}{K}_{ab} & = & -n_{c}n_{a}n_{b}\,\tilde{U}+2n_{c}n_{(a}\tilde{V}_{b)}+n_{a}n_{b}\tilde{W}_{c}\nonumber \\
	&  & -n_{c}\tilde{X}_{ab}-2\tilde{Y}_{c(a}n_{b)}+\tilde{Z}_{cab}, \label{nabla_Kt_dec_form}
	\end{eqnarray}
with
	\begin{eqnarray}
	\tilde{U} & = & \pounds_{\bm{n}}\overset{\mathrm{u}}{K}{}^{\perp\perp}-2a^{d}\overset{\mathrm{u}}{K}{}_{d}^{\perp\parallel},\\
	\tilde{V}_{b} & = & -a_{b}\overset{\mathrm{u}}{K}{}^{\perp\perp}+\pounds_{\bm{n}}\overset{\mathrm{u}}{K}{}_{b}^{\perp\parallel}-\overset{\mathrm{u}}{K}{}_{d}^{\perp\parallel}K_{b}^{d}-\overset{\mathrm{u}}{K}{}_{bd}^{\parallel\parallel}a^{d},\\
	\tilde{W}_{c} & = & \mathrm{D}_{c}\overset{\mathrm{u}}{K}{}^{\perp\perp}-2K_{c}^{d}\overset{\mathrm{u}}{K}{}_{d}^{\perp\parallel},\\
	\tilde{X}_{ab} & = & -2a_{(a}\overset{\mathrm{u}}{K}{}_{b)}^{\perp\parallel}+\pounds_{\bm{n}}\overset{\mathrm{u}}{K}{}_{ab}^{\parallel\parallel}-2\overset{\mathrm{u}}{K}{}_{d(a}^{\parallel\parallel}K_{b)}^{d},\\
	\tilde{Y}_{cb} & = & -K_{cb}\overset{\mathrm{u}}{K}{}^{\perp\perp}+\mathrm{D}_{c}\overset{\mathrm{u}}{K}{}_{b}^{\perp\parallel}-K_{c}^{d}\overset{\mathrm{u}}{K}{}_{db}^{\parallel\parallel},\\
	\tilde{Z}_{cab} & = & -2K_{c(a}\overset{\mathrm{u}}{K}{}_{b)}^{\perp\parallel}+\mathrm{D}_{c}\overset{\mathrm{u}}{K}{}_{ab}^{\parallel\parallel},
	\end{eqnarray}
where $\overset{\mathrm{u}}{K}{}^{\perp\perp}$, $\overset{\mathrm{u}}{K}{}_{a}^{\perp\parallel}$
and $\overset{\mathrm{u}}{K}{}_{ab}^{\parallel\parallel}$ are given in (\ref{Kab_Hind_nn}),
(\ref{Kab_Hind_nh}) and (\ref{Kab_Hind_hh}), respectively.
Plugging (\ref{nabla_Kt_dec_form}) into (\ref{cov_corr_DcKab}) (together with (\ref{ind_Hab_hab})), we are able to derive the full expansion of $\overset{\mathrm{u}}{\mathrm{D}}_{c}\overset{\mathrm{u}}{K}_{ab}$.
Similarly, 
	\begin{equation}
	\nabla_{a}\overset{\mathrm{u}}{a}_{b}=+n_{a}n_{b}\tilde{A}-n_{a}\tilde{B}_{b}-\tilde{C}_{a}n_{b}+\tilde{\Delta}_{ab}, \label{nabla_at_dec_form}
	\end{equation}
with
	\begin{eqnarray}
	\tilde{A} & = & \pounds_{\bm{n}}\overset{\mathrm{u}}{a}{}^{\perp}-a^{c}\overset{\mathrm{u}}{a}{}_{c}^{\parallel},\\
	\tilde{B}_{b} & = & -a_{b}\overset{\mathrm{u}}{a}{}^{\perp}+\pounds_{\bm{n}}\overset{\mathrm{u}}{a}{}_{b}^{\parallel}-K_{b}^{c}\overset{\mathrm{u}}{a}{}_{c}^{\parallel},\\
	\tilde{C}_{a} & = & \mathrm{D}_{a}\overset{\mathrm{u}}{a}{}^{\perp}-K_{a}^{c}\overset{\mathrm{u}}{a}{}_{c}^{\parallel},\\
	\tilde{\Delta}_{ab} & = & -K_{ab}\overset{\mathrm{u}}{a}{}^{\perp}+\mathrm{D}_{a}\overset{\mathrm{u}}{a}{}_{b}^{\parallel},
	\end{eqnarray}
where $\overset{\mathrm{u}}{a}{}^{\perp}$ and $\overset{\mathrm{u}}{a}{}_{a}^{\parallel}$ are given
in (\ref{acce_Hind_perp}) and (\ref{acce_Hind_parallel}), respectively.	
Plugging (\ref{nabla_at_dec_form}) into (\ref{cov_corr_Daab}) (together with (\ref{ind_Hab_hab})), we are able to derive the full expansion of $\overset{\mathrm{u}}{\mathrm{D}}_{a}\overset{\mathrm{u}}{a}_{b}$.

\section{Conclusion} \label{sec:con}

Significant achievements have been made in constructing ghostfree scalar-tensor theories which contain derivatives up to the second order.
However, theories with derivatives beyond the second order have not been well explored.
We make the first step in the current work, by connecting the spatially covariant gravity with higher derivative scalar-tensor theory, and using the former as ``generator'' of the latter with derivatives beyond the second order.

In Sec. \ref{sec:corr}, we present the explicit correspondences between the two types of theories. Precisely, we derive the maps between generally covariant derivatives of the scalar field and the spatially covariant geometric quantities such as the acceleration and the extrinsic curvature. 
The point is, the scalar-tensor terms that arise from the corresponding spatially covariant gravity are automatically ghostfree, at least as long as the scalar field possesses a timelike gradient.
In Sec. \ref{sec:cls}, we exhaust all the possible scalar monomials built of the extrinsic and intrinsic curvature $K_{ij}$ and $R_{ij}$ and the lapse function $N$ as well as their spatial derivative, up to the 4th order in the total number of derivatives.
We classify these monomials according to the total number of derivatives $d$, and further to the set of integers $(c_{0},c_{1},c_{2};d_{2},d_{3},d_{4})$ of the corresponding scalar-tensor monomials, which are developed in \cite{Gao:2020juc}.
This kind of classification not only captures the order of derivatives, but also makes the transferring from the spatially covariant gravity to the generally covariant higher derivative scalar-tensor theory transparent.

The correspondence discussed in Sec. \ref{sec:corr} and \ref{sec:cls} is valid only when the scalar field is timelike. 
In particular, the argument that the resulting scalar-tensor theory is ghostfree is only manifest in the unitary gauge.
In Sec. \ref{sec:covdec}, without fixing any gauge, we derive the 3+1 decomposition of the scalar-tensor terms that correspond to the acceleration and the extrinsic curvature as well as their spatial derivatives.
The results are given in (\ref{Kab_Hind_dec}), (\ref{acce_Hind_dec}), (\ref{cov_corr_DcKab}) and (\ref{cov_corr_Daab}).
From these results we can see that when deviating from the unitary gauge, $\pounds_{\bm{n}}\beta_{a}$ and $\pounds_{\bm{n}}^{2}\beta_{a}$ generally exist which apparently indicate the non-degeneracy of the theory.
These results will be used in studying the ghostfree scalar-tensor terms without assuming whether the scalar field is time or spacelike \cite{Gao:inprogress}.

\acknowledgments

This work was partly supported by the Natural Science Foundation of China (NSFC) under the grant No. 11975020.

\appendix

\section{Useful integrations by parts} \label{app:ibp}

Some of the scalar-tensor monomials are related to each other up to total derivatives, which can be used to reduce expressions at the level of Lagrangian through integrations by parts. 
Here we derive some useful results.

\subsection{$\nabla_{a}\nabla_{b}\phi\nabla^{a}\phi\nabla^{b}\phi$} \label{app:ibp_K}

For an arbitrary scalar function $f=f(\phi,X)$, by expanding the total derivative $\nabla_{a}\left(f\nabla^{a}\phi\right)$, we get
	\begin{equation}
	\frac{\partial f}{\partial X}\nabla_{a}\nabla_{b}\phi\nabla^{a}\phi\nabla^{b}\phi=f\square\phi-2X\frac{\partial f}{\partial\phi}-\nabla_{a}\left(f\nabla^{a}\phi\right), \label{ibp_100}
	\end{equation}
which gives the relation of $\nabla_{a}\nabla_{b}\phi\nabla^{a}\phi\nabla^{b}\phi$ and $\square\phi$ up to total derivatives.

(\ref{ibp_100}) can be used to simplify the scalar-tensor correspondence of $K$. For a general function $f=f(t,N)$, from (\ref{K_STxpl}) the scalar-tensor correspondence of $f\,K$ reads
	\begin{equation}
	f\,K\rightarrow-\frac{f}{\sqrt{2X}}\square\phi-\frac{f}{\left(2X\right)^{3/2}}\nabla_{a}\phi\nabla_{b}\phi\nabla^{a}\nabla^{b}\phi, \label{fK_ST}
	\end{equation}
where on the right-hand-side $f$ is understood to be $f=f(\phi,X)$.
By replacing $-\frac{f}{\left(2X\right)^{3/2}}\rightarrow\frac{\partial f}{\partial X}$ in (\ref{ibp_100}) and defining
	\begin{equation}
	F\left(\phi,X\right)=-\int^{X}\mathrm{d}Y\frac{f\left(\phi,Y\right)}{\left(2Y\right)^{3/2}},
	\end{equation}
(\ref{fK_ST}) becomes
	\begin{eqnarray}
	f\,K & \rightarrow & 2X\frac{\partial F}{\partial X}\square\phi-\left[-F\square\phi+2X\frac{\partial F}{\partial\phi}+\nabla_{a}\left(F\nabla^{a}\phi\right)\right]\nonumber \\
	& = & \left(F+2X\frac{\partial F}{\partial X}\right)\square\phi-2X\frac{\partial F}{\partial\phi}-\nabla_{a}\left(F\nabla^{a}\phi\right),
	\end{eqnarray}
which is thus (\ref{fK_ST_ibp}).

\subsection{${}^{4}\!R_{ab}\nabla^{a}\phi\nabla^{b}\phi$} \label{app:ibp_R3}

On a flat background, 
	\begin{equation}
	\left(\partial^{2}\phi\right)^2-\partial_{a}\partial_{b}\phi\partial^{a}\partial^{b}\phi\equiv\partial_{a}\left(\partial^{2}\phi\partial^{a}\phi-\partial_{b}\phi\partial^{a}\partial^{b}\phi\right),\label{td2_single}
	\end{equation}
which is a total derivative. Nevertheless, by replacing the ordinary
derivative by the covariant derivative directly and expanding the right-hand-side
of (\ref{td2_single}), we get
	\begin{eqnarray}
	&  & \nabla_{a}\left(\square\phi\nabla^{a}\phi-\nabla_{b}\phi\nabla^{a}\nabla^{b}\phi\right)\nonumber \\
	& = & \square\phi\square\phi-\nabla_{a}\nabla_{b}\phi\nabla^{a}\nabla^{b}\phi-\nabla^{a}\phi\left[\square,\nabla_{a}\right]\phi. \label{2nd_td}
	\end{eqnarray}
(\ref{2nd_td}) also implies that on a curved background $\square\phi\square\phi-\nabla_{a}\nabla_{b}\phi\nabla^{a}\nabla^{b}\phi$
is not a total derivative.
Instead, using the fact that the curvature tensor is the commutator
of two covariant derivatives, we get
	\begin{equation}
	{}^{4}\!R_{ab}\nabla^{a}\phi\nabla^{b}\phi=\left(\square\phi\right)^{2}-\nabla_{a}\nabla_{b}\phi\nabla^{a}\nabla^{b}\phi-\nabla_{a}\left(\square\phi\nabla^{a}\phi+\nabla^{a}X\right).\label{td2_curved}
	\end{equation}
Starting from (\ref{td2_curved}), we can derive a more general identity.
For an arbitrary function $F=F\left(\phi,X\right)$, after some manipulations and make use of (\ref{ibp_100}), we find
	\begin{eqnarray}
	&  & F\,{}^{4}\!R_{ab}\nabla^{a}\phi\nabla^{b}\phi\nonumber \\
	& = & F\left(\left(\square\phi\right)^{2}-\nabla_{a}\nabla_{b}\phi\nabla^{a}\nabla^{b}\phi\right)\nonumber \\
	&  & -\frac{\partial F}{\partial X}\left(\square\phi\nabla^{a}\phi\nabla^{b}\phi\nabla_{a}\nabla_{b}\phi-\nabla^{a}\phi\nabla^{b}\phi\nabla_{c}\nabla_{a}\phi\nabla^{c}\nabla_{b}\phi\right)\nonumber \\
	&  & +\left(\tilde{F}+2X\frac{\partial\tilde{F}}{\partial X}\right)\square\phi-2X\frac{\partial\tilde{F}}{\partial\phi}\nonumber \\
	&  & -\nabla_{a}\left[\tilde{F}\nabla^{a}\phi+F\left(\square\phi\nabla^{a}\phi+\nabla^{a}X\right)\right], \label{ibp_200}
	\end{eqnarray}
with
	\begin{equation}
		\tilde{F}\left(\phi,X\right)\equiv-\int^{X}\mathrm{d}Y\frac{\partial F\left(\phi,Y\right)}{\partial\phi}.
	\end{equation}
Obviously, (\ref{td2_curved}) is a special case of (\ref{ibp_200})
with $F=1$.
Similar to (\ref{ibp_100}), (\ref{ibp_200}) indicates that at the level of Lagrangian, $F\,\,{}^{4}\!R_{ab}\nabla^{a}\phi\nabla^{b}\phi$ is not independent, which can be expressed in terms of combinations of covariant derivatives of the scalar field only.

(\ref{ibp_200}) can be used to simplify the scalar-tensor correspondence of ${}^{3}\!R$. For a general function $f=f(t,N)$, from (\ref{R3_ST}) we get
	\begin{eqnarray}
	f\,{}^{3}\!R & \rightarrow & f\,{}^{4}\!R+\frac{f}{X}\,{}^{4}\!R_{ab}\nabla^{a}\phi\nabla^{b}\phi\nonumber \\
	&  & -\frac{f}{2X}\left(\square\phi\right)^{2}-\frac{f}{2X^{2}}\square\phi\nabla_{a}\phi\nabla_{b}\phi\nabla^{a}\nabla^{b}\phi\nonumber \\
	&  & +\frac{f}{2X}\nabla_{a}\nabla_{b}\phi\nabla^{a}\nabla^{b}\phi+\frac{f}{2X^{2}}\nabla^{a}\phi\nabla^{b}\phi\nabla_{c}\nabla_{a}\phi\nabla^{c}\nabla_{b}\phi,
	\end{eqnarray}
where again on the right-hand-side $f$ is understood as $f=f(\phi,X)$.
By replacing $F\rightarrow \frac{f}{X}$ in (\ref{ibp_200}), we get
	\begin{eqnarray}
	f\,{}^{3}\!R & \rightarrow & f\,{}^{4}\!R\nonumber \\
	&  & +\frac{f}{X}\left(\left(\square\phi\right)^{2}-\nabla_{a}\nabla_{b}\phi\nabla^{a}\nabla^{b}\phi\right)\nonumber \\
	&  & -\frac{\partial}{\partial X}\left(\frac{f}{X}\right)\left(\square\phi\nabla^{a}\phi\nabla^{b}\phi\nabla_{a}\nabla_{b}\phi-\nabla^{a}\phi\nabla^{b}\phi\nabla_{c}\nabla_{a}\phi\nabla^{c}\nabla_{b}\phi\right)\nonumber \\
	&  & +\left(F+2X\frac{\partial F}{\partial X}\right)\square\phi-2X\frac{\partial F}{\partial\phi}\nonumber \\
	&  & -\nabla_{a}\left[F\nabla^{a}\phi+\frac{f}{X}\left(\square\phi\nabla^{a}\phi+\nabla^{a}X\right)\right],\nonumber \\
	&  & -\frac{f}{2X}\left(\square\phi\right)^{2}-\frac{f}{2X^{2}}\square\phi\nabla_{a}\phi\nabla_{b}\phi\nabla^{a}\nabla^{b}\phi\nonumber \\
	&  & +\frac{f}{2X}\nabla_{a}\nabla_{b}\phi\nabla^{a}\nabla^{b}\phi+\frac{f}{2X^{2}}\nabla^{a}\phi\nabla^{b}\phi\nabla_{c}\nabla_{a}\phi\nabla^{c}\nabla_{b}\phi,
	\end{eqnarray}
with
	\begin{equation}
	F\left(\phi,X\right)\equiv-\int^{X}\mathrm{d}Y\frac{1}{Y}\frac{\partial f\left(\phi,Y\right)}{\partial\phi}.
	\end{equation}
After some simplifications, we thus arrive at (\ref{R3_STibp}).

\section{Explicit expressions for the decomposition of $\nabla_{c}\nabla_{a}\nabla_{b}\phi$} \label{app:xpld3sf}

The decomposition of $\nabla_{c}\nabla_{a}\nabla_{b}\phi$ can always be written in the form (\ref{nabla3_sca_dec_form}). After plugging (\ref{nabla2_sca_dec_A})-(\ref{nabla2_sca_dec_Delta}) into (\ref{nabla3_sca_dec_form}), we find
	\begin{eqnarray}
	U & = & \pounds_{\bm{n}}^{3}\phi+2a^{d}a_{d}\pounds_{\bm{n}}\phi-\pounds_{\bm{n}}\left(a^{d}\mathrm{D}_{d}\phi\right)\nonumber \\
	&  & -2a^{d}\pounds_{\bm{n}}\mathrm{D}_{d}\phi+2a^{d}K_{d}^{e}\mathrm{D}_{e}\phi,\label{nabla3_sca_dec_U_xpl}
	\end{eqnarray}
	\begin{eqnarray}
	V_{b} & = & -2a_{b}\pounds_{\bm{n}}^{2}\phi-\left(\pounds_{\bm{n}}a_{b}-2a_{d}K_{b}^{d}\right)\pounds_{\bm{n}}\phi\nonumber \\
	&  & +\pounds_{\bm{n}}^{2}\mathrm{D}_{b}\phi-\pounds_{\bm{n}}\left(K_{b}^{d}\mathrm{D}_{d}\phi\right)-K_{b}^{d}\pounds_{\bm{n}}\mathrm{D}_{d}\phi\nonumber \\
	&  & +a_{b}a^{d}\mathrm{D}_{d}\phi+K_{b}^{d}K_{d}^{e}\mathrm{D}_{e}\phi-a^{d}\mathrm{D}_{d}\mathrm{D}_{b}\phi,\label{nabla3_sca_dec_V_xpl}
	\end{eqnarray}
	\begin{eqnarray}
	W_{c} & = & -2a_{c}\pounds_{\bm{n}}^{2}\phi-\left(\pounds_{\bm{n}}a_{c}-2a_{d}K_{c}^{d}\right)\pounds_{\bm{n}}\phi+\pounds_{\bm{n}}^{2}\mathrm{D}_{c}\phi\nonumber \\
	&  & -2K_{c}^{d}\pounds_{\bm{n}}\mathrm{D}_{d}\phi-\mathrm{D}_{c}\left(a^{d}\mathrm{D}_{d}\phi\right)+2K_{c}^{d}K_{d}^{e}\mathrm{D}_{e}\phi, \label{nabla3_sca_dec_W_xpl}
	\end{eqnarray}
	\begin{eqnarray}
	X_{ab} & = & -K_{ab}\pounds_{\bm{n}}^{2}\phi+\left(2a_{a}a_{b}+2K_{a}^{d}K_{bd}-\pounds_{\bm{n}}K_{ab}\right)\pounds_{\bm{n}}\phi\nonumber \\
	&  & -2a_{(a}\pounds_{\bm{n}}\mathrm{D}_{b)}\phi+\pounds_{\bm{n}}\left(\mathrm{D}_{a}\mathrm{D}_{b}\phi\right)\nonumber \\
	&  & +2a_{(a}K_{b)}^{d}\mathrm{D}_{d}\phi-2K_{(a}^{d}\mathrm{D}_{b)}\mathrm{D}_{d}\phi,\label{nabla3_sca_dec_X_xpl}
	\end{eqnarray}
	\begin{eqnarray}
	Y_{cb} & = & -K_{cb}\pounds_{\bm{n}}^{2}\phi+\left(K_{c}^{d}K_{db}-\mathrm{D}_{c}a_{b}+a_{c}a_{b}\right)\pounds_{\bm{n}}\phi\nonumber \\
	&  & -a_{b}\pounds_{\bm{n}}\mathrm{D}_{c}\phi+\mathrm{D}_{c}\left(\pounds_{\bm{n}}\mathrm{D}_{b}\phi\right)-\mathrm{D}_{c}K_{b}^{d}\mathrm{D}_{d}\phi\nonumber \\
	&  & +K_{cb}a^{d}\mathrm{D}_{d}\phi-2K_{(c}^{d}\mathrm{D}_{b)}\mathrm{D}_{d}\phi,\label{nabla3_sca_dec_Y_xpl}
	\end{eqnarray}
and
	\begin{eqnarray}
	Z_{cab} & = & \left(-\mathrm{D}_{c}K_{ab}+3a_{(c}K_{ab)}\right)\pounds_{\bm{n}}\phi-3K_{(ca}\pounds_{\bm{n}}\mathrm{D}_{b)}\phi\nonumber \\
	&  & +2K_{c(a}K_{b)}^{d}\mathrm{D}_{d}\phi+\mathrm{D}_{c}\mathrm{D}_{a}\mathrm{D}_{b}\phi.\label{nabla3_sca_dec_Z_xpl}
	\end{eqnarray}

\section{Expressions of scalar-tensor monomials} \label{app:STmono}

The scalar-tensor monomials are investigated in \cite{Gao:2020juc}. Here we collect the expressions for the unfactorizable monomials used in this work. We refer to \cite{Gao:2020juc} for the detailed discussion.

\subsection{$d=1$}

We define
	\begin{eqnarray}
	\bm{E}_{1}^{\left(0;1,0\right)} & \equiv & \frac{1}{\sigma}\square\phi, \label{E010_1}\\
	\bm{E}_{2}^{\left(0;1,0\right)} & \equiv & \frac{1}{\sigma^{3}}\nabla_{a}\phi\nabla_{b}\phi\nabla^{a}\nabla^{b}\phi, \label{E010_2}
	\end{eqnarray}
where and in what follows we denote
\begin{equation}
\sigma = \sqrt{2X}
\end{equation}
for short.

\subsection{$d=2$}

We define
\begin{eqnarray}
\bm{E}_{1}^{\left(0;2,0\right)} & \equiv & \frac{1}{\sigma^{2}}\nabla_{a}\nabla_{b}\phi\nabla^{a}\nabla^{b}\phi,\\
\bm{E}_{2}^{\left(0;2,0\right)} & \equiv & \frac{1}{\sigma^{4}}\nabla^{a}\phi\nabla^{b}\phi\nabla_{c}\nabla_{a}\phi\nabla^{c}\nabla_{b}\phi,
\end{eqnarray}
\begin{eqnarray}
\bm{E}_{1}^{\left(0;0,1\right)} & \equiv & \frac{1}{\sigma^{2}}\nabla^{a}\phi\nabla_{a}\square\phi,\\
\bm{E}_{2}^{\left(0;0,1\right)} & \equiv & \frac{1}{\sigma^{2}}\nabla^{a}\phi\square\nabla_{a}\phi,\\
\bm{E}_{3}^{\left(0;0,1\right)} & \equiv & \frac{1}{\sigma^{4}}\nabla^{a}\phi\nabla^{b}\phi\nabla^{c}\phi\nabla_{a}\nabla_{b}\nabla_{c}\phi,
\end{eqnarray}
and
\begin{eqnarray}
\bm{E}_{1}^{\left(1;0,0\right)} & \equiv & R,\\
\bm{E}_{2}^{\left(1;0,0\right)} & \equiv & \frac{1}{\sigma^{2}}R_{ab}\nabla^{a}\phi\nabla^{b}\phi.
\end{eqnarray}

\subsection{$d=3$}

We define
\begin{eqnarray}
\bm{E}_{1}^{\left(0;3,0\right)} & \equiv & \frac{1}{\sigma^{3}}\nabla_{a}\nabla^{b}\phi\nabla_{b}\nabla^{c}\phi\nabla_{c}\nabla^{a}\phi,\\
\bm{E}_{2}^{\left(0;3,0\right)} & \equiv & \frac{1}{\sigma^{5}}\nabla^{a}\phi\nabla^{b}\phi\nabla_{a}\nabla_{c}\phi\nabla^{c}\nabla_{d}\phi\nabla^{d}\nabla_{b}\phi,
\end{eqnarray}
\begin{eqnarray}
\bm{E}_{1}^{\left(0;1,1\right)} & \equiv & \frac{1}{\sigma^{3}}\nabla^{a}\phi\nabla_{a}\nabla^{b}\phi\nabla_{b}\square\phi,\\
\bm{E}_{2}^{\left(0;1,1\right)} & \equiv & \frac{1}{\sigma^{3}}\nabla^{a}\phi\nabla_{a}\nabla^{b}\phi\square\nabla_{b}\phi,\\
\bm{E}_{3}^{\left(0;1,1\right)} & \equiv & \frac{1}{\sigma^{3}}\nabla^{a}\phi\nabla^{b}\nabla^{c}\phi\nabla_{a}\nabla_{b}\nabla_{c}\phi,\\
\bm{E}_{4}^{\left(0;1,1\right)} & \equiv & \frac{1}{\sigma^{3}}\nabla^{a}\phi\nabla^{b}\nabla^{c}\phi\nabla_{b}\nabla_{c}\nabla_{a}\phi,\\
\bm{E}_{5}^{\left(0;1,1\right)} & \equiv & \frac{1}{\sigma^{5}}\nabla^{a}\phi\nabla^{b}\phi\nabla^{c}\phi\nabla_{a}\nabla^{d}\phi\nabla_{d}\nabla_{b}\nabla_{c}\phi,
\end{eqnarray}
\begin{eqnarray}
\bm{E}_{1}^{\left(1;1,0\right)} & \equiv & \frac{1}{\sigma}R_{ab}\nabla^{a}\nabla^{b}\phi,\\
\bm{E}_{2}^{\left(1;1,0\right)} & \equiv & \frac{1}{\sigma^{3}}R_{abcd}\nabla^{a}\phi\nabla^{c}\phi\nabla^{b}\nabla^{d}\phi,\\
\bm{E}_{3}^{\left(1;1,0\right)} & \equiv & \frac{1}{\sigma^{3}}R_{ab}\nabla^{a}\phi\nabla^{c}\phi\nabla^{b}\nabla_{c}\phi.
\end{eqnarray}
In the case of parity violation, we define
	\begin{equation}
	\bm{F}_{1}^{\left(0;1,1\right)} \equiv \frac{1}{\sigma^{3}}\varepsilon_{abcd}\nabla^{a}\phi\nabla^{b}\nabla^{f}\phi\nabla^{c}\nabla^{d}\nabla_{f}\phi,
	\end{equation}
and
	\begin{equation}
	\bm{F}_{1}^{\left(1;1,0\right)}\equiv\frac{1}{\sigma^{3}}\varepsilon_{abcd}R_{ef}^{\phantom{ef}cd}\nabla^{a}\phi\nabla^{e}\phi\nabla^{b}\nabla^{f}\phi,
	\end{equation}
Note
	\begin{equation}
	\bm{F}_{1}^{\left(0;1,1\right)}\equiv-\frac{1}{2}\bm{F}_{1}^{(1;1,0)}. \label{F011_F110}
	\end{equation}

\subsection{Factorizable monomials with $d=4$} \label{app:d4fac}

The scalar-tensor correspondence of the factorizable monomials can be read easily. Here we show their expressions for completeness:
\begin{equation}
KK_{ij}K^{ij}\rightarrow-\left(\bm{E}_{1}^{(0;1,0)}+\bm{E}_{2}^{(0;1,0)}\right)\left(\left(\bm{E}_{2}^{(0;1,0)}\right)^{2}+\bm{E}_{1}^{(0;2,0)}+2\bm{E}_{2}^{(0;2,0)}\right),\label{KKabKab_STM}
\end{equation}
\begin{equation}
K^{3}=-\left(\bm{E}_{1}^{(0;1,0)}+\bm{E}_{2}^{(0;1,0)}\right)^{3},\label{KKK_STm}
\end{equation}
\begin{equation}
Ka_{i}a^{i}\rightarrow-\left(\bm{E}_{1}^{(0;1,0)}+\bm{E}_{2}^{(0;1,0)}\right)\left(\left(\bm{E}_{2}^{(0;1,0)}\right)^{2}+\bm{E}_{2}^{(0;2,0)}\right),\label{Kaaaa_STm}
\end{equation}
\begin{eqnarray}
^{3}\!RK & \rightarrow & -\left(\bm{E}_{1}^{(0;1,0)}+\bm{E}_{2}^{(0;1,0)}\right)\nonumber \\
&  & \times\left(\,{}^{4}\!R+2\bm{E}_{2}^{(1;0,0)}-\left(\bm{E}_{1}^{(0;1,0)}\right)^{2}-2\bm{E}_{1}^{(0;1,0)}\bm{E}_{2}^{(0;1,0)}+\bm{E}_{1}^{(0;2,0)}+2\bm{E}_{2}^{(0;2,0)}\right),\label{R3K_STm}
\end{eqnarray}
and
\begin{eqnarray}
K\nabla_{i}a^{i} & \rightarrow & -\left(\bm{E}_{1}^{(0;1,0)}+\bm{E}_{2}^{(0;1,0)}\right)\nonumber \\
&  & \times\left(\bm{E}_{2}^{(0;0,1)}+\bm{E}_{3}^{(0;0,1)}+\bm{E}_{1}^{(0;1,0)}\bm{E}_{2}^{(0;1,0)}+3\left(\bm{E}_{2}^{(0;1,0)}\right)^{2}+\bm{E}_{1}^{(0;2,0)}+3\bm{E}_{2}^{(0;2,0)}\right).\label{KDaaa_STm}
\end{eqnarray}

%


\begin{thebibliography}{51}%
	\makeatletter
	\providecommand \@ifxundefined [1]{%
		\@ifx{#1\undefined}
	}%
	\providecommand \@ifnum [1]{%
		\ifnum #1\expandafter \@firstoftwo
		\else \expandafter \@secondoftwo
		\fi
	}%
	\providecommand \@ifx [1]{%
		\ifx #1\expandafter \@firstoftwo
		\else \expandafter \@secondoftwo
		\fi
	}%
	\providecommand \natexlab [1]{#1}%
	\providecommand \enquote  [1]{``#1''}%
	\providecommand \bibnamefont  [1]{#1}%
	\providecommand \bibfnamefont [1]{#1}%
	\providecommand \citenamefont [1]{#1}%
	\providecommand \href@noop [0]{\@secondoftwo}%
	\providecommand \href [0]{\begingroup \@sanitize@url \@href}%
	\providecommand \@href[1]{\@@startlink{#1}\@@href}%
	\providecommand \@@href[1]{\endgroup#1\@@endlink}%
	\providecommand \@sanitize@url [0]{\catcode `\\12\catcode `\$12\catcode
		`\&12\catcode `\#12\catcode `\^12\catcode `\_12\catcode `\%12\relax}%
	\providecommand \@@startlink[1]{}%
	\providecommand \@@endlink[0]{}%
	\providecommand \url  [0]{\begingroup\@sanitize@url \@url }%
	\providecommand \@url [1]{\endgroup\@href {#1}{\urlprefix }}%
	\providecommand \urlprefix  [0]{URL }%
	\providecommand \Eprint [0]{\href }%
	\providecommand \doibase [0]{http://dx.doi.org/}%
	\providecommand \selectlanguage [0]{\@gobble}%
	\providecommand \bibinfo  [0]{\@secondoftwo}%
	\providecommand \bibfield  [0]{\@secondoftwo}%
	\providecommand \translation [1]{[#1]}%
	\providecommand \BibitemOpen [0]{}%
	\providecommand \bibitemStop [0]{}%
	\providecommand \bibitemNoStop [0]{.\EOS\space}%
	\providecommand \EOS [0]{\spacefactor3000\relax}%
	\providecommand \BibitemShut  [1]{\csname bibitem#1\endcsname}%
	\let\auto@bib@innerbib\@empty
	\bibitem [{\citenamefont {Woodard}(2015)}]{Woodard:2015zca}%
	\BibitemOpen
	\bibfield  {author} {\bibinfo {author} {\bibfnamefont {R.~P.}\ \bibnamefont
			{Woodard}},\ }\href {\doibase 10.4249/scholarpedia.32243} {\bibfield
		{journal} {\bibinfo  {journal} {Scholarpedia}\ }\textbf {\bibinfo {volume}
			{10}},\ \bibinfo {pages} {32243} (\bibinfo {year} {2015})},\ \Eprint
	{http://arxiv.org/abs/1506.02210} {arXiv:1506.02210 [hep-th]} \BibitemShut
	{NoStop}%
	\bibitem [{\citenamefont {Horndeski}(1974)}]{Horndeski:1974wa}%
	\BibitemOpen
	\bibfield  {author} {\bibinfo {author} {\bibfnamefont {G.~W.}\ \bibnamefont
			{Horndeski}},\ }\href {\doibase 10.1007/BF01807638} {\bibfield  {journal}
		{\bibinfo  {journal} {Int.J.Theor.Phys.}\ }\textbf {\bibinfo {volume} {10}},\
		\bibinfo {pages} {363} (\bibinfo {year} {1974})}\BibitemShut {NoStop}%
	\bibitem [{\citenamefont {Nicolis}\ \emph {et~al.}(2009)\citenamefont
		{Nicolis}, \citenamefont {Rattazzi},\ and\ \citenamefont
		{Trincherini}}]{Nicolis:2008in}%
	\BibitemOpen
	\bibfield  {author} {\bibinfo {author} {\bibfnamefont {A.}~\bibnamefont
			{Nicolis}}, \bibinfo {author} {\bibfnamefont {R.}~\bibnamefont {Rattazzi}}, \
		and\ \bibinfo {author} {\bibfnamefont {E.}~\bibnamefont {Trincherini}},\
	}\href {\doibase 10.1103/PhysRevD.79.064036} {\bibfield  {journal} {\bibinfo
			{journal} {Phys.Rev.}\ }\textbf {\bibinfo {volume} {D79}},\ \bibinfo {pages}
		{064036} (\bibinfo {year} {2009})},\ \Eprint {http://arxiv.org/abs/0811.2197}
	{arXiv:0811.2197 [hep-th]} \BibitemShut {NoStop}%
	\bibitem [{\citenamefont {Deffayet}\ \emph {et~al.}(2011)\citenamefont
		{Deffayet}, \citenamefont {Gao}, \citenamefont {Steer},\ and\ \citenamefont
		{Zahariade}}]{Deffayet:2011gz}%
	\BibitemOpen
	\bibfield  {author} {\bibinfo {author} {\bibfnamefont {C.}~\bibnamefont
			{Deffayet}}, \bibinfo {author} {\bibfnamefont {X.}~\bibnamefont {Gao}},
		\bibinfo {author} {\bibfnamefont {D.}~\bibnamefont {Steer}}, \ and\ \bibinfo
		{author} {\bibfnamefont {G.}~\bibnamefont {Zahariade}},\ }\href {\doibase
		10.1103/PhysRevD.84.064039} {\bibfield  {journal} {\bibinfo  {journal}
			{Phys.Rev.}\ }\textbf {\bibinfo {volume} {D84}},\ \bibinfo {pages} {064039}
		(\bibinfo {year} {2011})},\ \Eprint {http://arxiv.org/abs/1103.3260}
	{arXiv:1103.3260 [hep-th]} \BibitemShut {NoStop}%
	\bibitem [{\citenamefont {Kobayashi}\ \emph {et~al.}(2011)\citenamefont
		{Kobayashi}, \citenamefont {Yamaguchi},\ and\ \citenamefont
		{Yokoyama}}]{Kobayashi:2011nu}%
	\BibitemOpen
	\bibfield  {author} {\bibinfo {author} {\bibfnamefont {T.}~\bibnamefont
			{Kobayashi}}, \bibinfo {author} {\bibfnamefont {M.}~\bibnamefont
			{Yamaguchi}}, \ and\ \bibinfo {author} {\bibfnamefont {J.}~\bibnamefont
			{Yokoyama}},\ }\href {\doibase 10.1143/PTP.126.511} {\bibfield  {journal}
		{\bibinfo  {journal} {Prog.Theor.Phys.}\ }\textbf {\bibinfo {volume} {126}},\
		\bibinfo {pages} {511} (\bibinfo {year} {2011})},\ \Eprint
	{http://arxiv.org/abs/1105.5723} {arXiv:1105.5723 [hep-th]} \BibitemShut
	{NoStop}%
	\bibitem [{\citenamefont {Gleyzes}\ \emph
		{et~al.}(2015{\natexlab{a}})\citenamefont {Gleyzes}, \citenamefont
		{Langlois}, \citenamefont {Piazza},\ and\ \citenamefont
		{Vernizzi}}]{Gleyzes:2014dya}%
	\BibitemOpen
	\bibfield  {author} {\bibinfo {author} {\bibfnamefont {J.}~\bibnamefont
			{Gleyzes}}, \bibinfo {author} {\bibfnamefont {D.}~\bibnamefont {Langlois}},
		\bibinfo {author} {\bibfnamefont {F.}~\bibnamefont {Piazza}}, \ and\ \bibinfo
		{author} {\bibfnamefont {F.}~\bibnamefont {Vernizzi}},\ }\href {\doibase
		10.1103/PhysRevLett.114.211101} {\bibfield  {journal} {\bibinfo  {journal}
			{Phys. Rev. Lett.}\ }\textbf {\bibinfo {volume} {114}},\ \bibinfo {pages}
		{211101} (\bibinfo {year} {2015}{\natexlab{a}})},\ \Eprint
	{http://arxiv.org/abs/1404.6495} {arXiv:1404.6495 [hep-th]} \BibitemShut
	{NoStop}%
	\bibitem [{\citenamefont {Gleyzes}\ \emph
		{et~al.}(2015{\natexlab{b}})\citenamefont {Gleyzes}, \citenamefont
		{Langlois}, \citenamefont {Piazza},\ and\ \citenamefont
		{Vernizzi}}]{Gleyzes:2014qga}%
	\BibitemOpen
	\bibfield  {author} {\bibinfo {author} {\bibfnamefont {J.}~\bibnamefont
			{Gleyzes}}, \bibinfo {author} {\bibfnamefont {D.}~\bibnamefont {Langlois}},
		\bibinfo {author} {\bibfnamefont {F.}~\bibnamefont {Piazza}}, \ and\ \bibinfo
		{author} {\bibfnamefont {F.}~\bibnamefont {Vernizzi}},\ }\href {\doibase
		10.1088/1475-7516/2015/02/018} {\bibfield  {journal} {\bibinfo  {journal}
			{JCAP}\ }\textbf {\bibinfo {volume} {1502}},\ \bibinfo {pages} {018}
		(\bibinfo {year} {2015}{\natexlab{b}})},\ \Eprint
	{http://arxiv.org/abs/1408.1952} {arXiv:1408.1952 [astro-ph.CO]} \BibitemShut
	{NoStop}%
	\bibitem [{\citenamefont {Langlois}\ and\ \citenamefont
		{Noui}(2016)}]{Langlois:2015cwa}%
	\BibitemOpen
	\bibfield  {author} {\bibinfo {author} {\bibfnamefont {D.}~\bibnamefont
			{Langlois}}\ and\ \bibinfo {author} {\bibfnamefont {K.}~\bibnamefont
			{Noui}},\ }\href {\doibase 10.1088/1475-7516/2016/02/034} {\bibfield
		{journal} {\bibinfo  {journal} {JCAP}\ }\textbf {\bibinfo {volume} {1602}},\
		\bibinfo {pages} {034} (\bibinfo {year} {2016})},\ \Eprint
	{http://arxiv.org/abs/1510.06930} {arXiv:1510.06930 [gr-qc]} \BibitemShut
	{NoStop}%
	\bibitem [{\citenamefont {Motohashi}\ \emph {et~al.}(2016)\citenamefont
		{Motohashi}, \citenamefont {Noui}, \citenamefont {Suyama}, \citenamefont
		{Yamaguchi},\ and\ \citenamefont {Langlois}}]{Motohashi:2016ftl}%
	\BibitemOpen
	\bibfield  {author} {\bibinfo {author} {\bibfnamefont {H.}~\bibnamefont
			{Motohashi}}, \bibinfo {author} {\bibfnamefont {K.}~\bibnamefont {Noui}},
		\bibinfo {author} {\bibfnamefont {T.}~\bibnamefont {Suyama}}, \bibinfo
		{author} {\bibfnamefont {M.}~\bibnamefont {Yamaguchi}}, \ and\ \bibinfo
		{author} {\bibfnamefont {D.}~\bibnamefont {Langlois}},\ }\href {\doibase
		10.1088/1475-7516/2016/07/033} {\bibfield  {journal} {\bibinfo  {journal}
			{JCAP}\ }\textbf {\bibinfo {volume} {1607}},\ \bibinfo {pages} {033}
		(\bibinfo {year} {2016})},\ \Eprint {http://arxiv.org/abs/1603.09355}
	{arXiv:1603.09355 [hep-th]} \BibitemShut {NoStop}%
	\bibitem [{\citenamefont {Langlois}(2019)}]{Langlois:2018dxi}%
	\BibitemOpen
	\bibfield  {author} {\bibinfo {author} {\bibfnamefont {D.}~\bibnamefont
			{Langlois}},\ }\href {\doibase 10.1142/S0218271819420069} {\bibfield
		{journal} {\bibinfo  {journal} {Int. J. Mod. Phys.}\ }\textbf {\bibinfo
			{volume} {D28}},\ \bibinfo {pages} {1942006} (\bibinfo {year} {2019})},\
	\Eprint {http://arxiv.org/abs/1811.06271} {arXiv:1811.06271 [gr-qc]}
	\BibitemShut {NoStop}%
	\bibitem [{\citenamefont {Kobayashi}(2019)}]{Kobayashi:2019hrl}%
	\BibitemOpen
	\bibfield  {author} {\bibinfo {author} {\bibfnamefont {T.}~\bibnamefont
			{Kobayashi}},\ }\href {\doibase 10.1088/1361-6633/ab2429} {\bibfield
		{journal} {\bibinfo  {journal} {Rept. Prog. Phys.}\ }\textbf {\bibinfo
			{volume} {82}},\ \bibinfo {pages} {086901} (\bibinfo {year} {2019})},\
	\Eprint {http://arxiv.org/abs/1901.07183} {arXiv:1901.07183 [gr-qc]}
	\BibitemShut {NoStop}%
	\bibitem [{\citenamefont {Buoninfante}\ \emph {et~al.}(2019)\citenamefont
		{Buoninfante}, \citenamefont {Lambiase},\ and\ \citenamefont
		{Yamaguchi}}]{Buoninfante:2018lnh}%
	\BibitemOpen
	\bibfield  {author} {\bibinfo {author} {\bibfnamefont {L.}~\bibnamefont
			{Buoninfante}}, \bibinfo {author} {\bibfnamefont {G.}~\bibnamefont
			{Lambiase}}, \ and\ \bibinfo {author} {\bibfnamefont {M.}~\bibnamefont
			{Yamaguchi}},\ }\href {\doibase 10.1103/PhysRevD.100.026019} {\bibfield
		{journal} {\bibinfo  {journal} {Phys. Rev.}\ }\textbf {\bibinfo {volume}
			{D100}},\ \bibinfo {pages} {026019} (\bibinfo {year} {2019})},\ \Eprint
	{http://arxiv.org/abs/1812.10105} {arXiv:1812.10105 [hep-th]} \BibitemShut
	{NoStop}%
	\bibitem [{\citenamefont {Buoninfante}\ \emph {et~al.}(2020)\citenamefont
		{Buoninfante}, \citenamefont {Lambiase}, \citenamefont {Miyashita},
		\citenamefont {Takebe},\ and\ \citenamefont
		{Yamaguchi}}]{Buoninfante:2020ctr}%
	\BibitemOpen
	\bibfield  {author} {\bibinfo {author} {\bibfnamefont {L.}~\bibnamefont
			{Buoninfante}}, \bibinfo {author} {\bibfnamefont {G.}~\bibnamefont
			{Lambiase}}, \bibinfo {author} {\bibfnamefont {Y.}~\bibnamefont {Miyashita}},
		\bibinfo {author} {\bibfnamefont {W.}~\bibnamefont {Takebe}}, \ and\ \bibinfo
		{author} {\bibfnamefont {M.}~\bibnamefont {Yamaguchi}},\ }\href {\doibase
		10.1103/PhysRevD.101.084019} {\bibfield  {journal} {\bibinfo  {journal}
			{Phys.\ Rev.\ D}\ }\textbf {\bibinfo {volume} {101}},\ \bibinfo {pages}
		{084019} (\bibinfo {year} {2020})},\ \Eprint
	{http://arxiv.org/abs/2001.07830} {arXiv:2001.07830 [hep-th]} \BibitemShut
	{NoStop}%
	\bibitem [{\citenamefont {Gao}(2020)}]{Gao:2020juc}%
	\BibitemOpen
	\bibfield  {author} {\bibinfo {author} {\bibfnamefont {X.}~\bibnamefont
			{Gao}},\ }\href@noop {} {\  (\bibinfo {year} {2020})},\ \Eprint
	{http://arxiv.org/abs/2003.11978} {arXiv:2003.11978 [gr-qc]} \BibitemShut
	{NoStop}%
	\bibitem [{\citenamefont {Lue}\ \emph {et~al.}(1999)\citenamefont {Lue},
		\citenamefont {Wang},\ and\ \citenamefont {Kamionkowski}}]{Lue:1998mq}%
	\BibitemOpen
	\bibfield  {author} {\bibinfo {author} {\bibfnamefont {A.}~\bibnamefont
			{Lue}}, \bibinfo {author} {\bibfnamefont {L.-M.}\ \bibnamefont {Wang}}, \
		and\ \bibinfo {author} {\bibfnamefont {M.}~\bibnamefont {Kamionkowski}},\
	}\href {\doibase 10.1103/PhysRevLett.83.1506} {\bibfield  {journal} {\bibinfo
			{journal} {Phys. Rev. Lett.}\ }\textbf {\bibinfo {volume} {83}},\ \bibinfo
		{pages} {1506} (\bibinfo {year} {1999})},\ \Eprint
	{http://arxiv.org/abs/astro-ph/9812088} {arXiv:astro-ph/9812088 [astro-ph]}
	\BibitemShut {NoStop}%
	\bibitem [{\citenamefont {Jackiw}\ and\ \citenamefont
		{Pi}(2003)}]{Jackiw:2003pm}%
	\BibitemOpen
	\bibfield  {author} {\bibinfo {author} {\bibfnamefont {R.}~\bibnamefont
			{Jackiw}}\ and\ \bibinfo {author} {\bibfnamefont {S.~Y.}\ \bibnamefont
			{Pi}},\ }\href {\doibase 10.1103/PhysRevD.68.104012} {\bibfield  {journal}
		{\bibinfo  {journal} {Phys. Rev.}\ }\textbf {\bibinfo {volume} {D68}},\
		\bibinfo {pages} {104012} (\bibinfo {year} {2003})},\ \Eprint
	{http://arxiv.org/abs/gr-qc/0308071} {arXiv:gr-qc/0308071 [gr-qc]}
	\BibitemShut {NoStop}%
	\bibitem [{\citenamefont {Deruelle}\ \emph {et~al.}(2012)\citenamefont
		{Deruelle}, \citenamefont {Sasaki}, \citenamefont {Sendouda},\ and\
		\citenamefont {Youssef}}]{Deruelle:2012xv}%
	\BibitemOpen
	\bibfield  {author} {\bibinfo {author} {\bibfnamefont {N.}~\bibnamefont
			{Deruelle}}, \bibinfo {author} {\bibfnamefont {M.}~\bibnamefont {Sasaki}},
		\bibinfo {author} {\bibfnamefont {Y.}~\bibnamefont {Sendouda}}, \ and\
		\bibinfo {author} {\bibfnamefont {A.}~\bibnamefont {Youssef}},\ }\href
	{\doibase 10.1007/JHEP09(2012)009} {\bibfield  {journal} {\bibinfo  {journal}
			{JHEP}\ }\textbf {\bibinfo {volume} {09}},\ \bibinfo {pages} {009} (\bibinfo
		{year} {2012})},\ \Eprint {http://arxiv.org/abs/1202.3131} {arXiv:1202.3131
		[gr-qc]} \BibitemShut {NoStop}%
	\bibitem [{\citenamefont {Crisostomi}\ \emph {et~al.}(2018)\citenamefont
		{Crisostomi}, \citenamefont {Noui}, \citenamefont {Charmousis},\ and\
		\citenamefont {Langlois}}]{Crisostomi:2017ugk}%
	\BibitemOpen
	\bibfield  {author} {\bibinfo {author} {\bibfnamefont {M.}~\bibnamefont
			{Crisostomi}}, \bibinfo {author} {\bibfnamefont {K.}~\bibnamefont {Noui}},
		\bibinfo {author} {\bibfnamefont {C.}~\bibnamefont {Charmousis}}, \ and\
		\bibinfo {author} {\bibfnamefont {D.}~\bibnamefont {Langlois}},\ }\href
	{\doibase 10.1103/PhysRevD.97.044034} {\bibfield  {journal} {\bibinfo
			{journal} {Phys. Rev.}\ }\textbf {\bibinfo {volume} {D97}},\ \bibinfo {pages}
		{044034} (\bibinfo {year} {2018})},\ \Eprint
	{http://arxiv.org/abs/1710.04531} {arXiv:1710.04531 [hep-th]} \BibitemShut
	{NoStop}%
	\bibitem [{\citenamefont {Creminelli}\ and\ \citenamefont
		{Vernizzi}(2017)}]{Creminelli:2017sry}%
	\BibitemOpen
	\bibfield  {author} {\bibinfo {author} {\bibfnamefont {P.}~\bibnamefont
			{Creminelli}}\ and\ \bibinfo {author} {\bibfnamefont {F.}~\bibnamefont
			{Vernizzi}},\ }\href {\doibase 10.1103/PhysRevLett.119.251302} {\bibfield
		{journal} {\bibinfo  {journal} {Phys. Rev. Lett.}\ }\textbf {\bibinfo
			{volume} {119}},\ \bibinfo {pages} {251302} (\bibinfo {year} {2017})},\
	\Eprint {http://arxiv.org/abs/1710.05877} {arXiv:1710.05877 [astro-ph.CO]}
	\BibitemShut {NoStop}%
	\bibitem [{\citenamefont {Sakstein}\ and\ \citenamefont
		{Jain}(2017)}]{Sakstein:2017xjx}%
	\BibitemOpen
	\bibfield  {author} {\bibinfo {author} {\bibfnamefont {J.}~\bibnamefont
			{Sakstein}}\ and\ \bibinfo {author} {\bibfnamefont {B.}~\bibnamefont
			{Jain}},\ }\href {\doibase 10.1103/PhysRevLett.119.251303} {\bibfield
		{journal} {\bibinfo  {journal} {Phys. Rev. Lett.}\ }\textbf {\bibinfo
			{volume} {119}},\ \bibinfo {pages} {251303} (\bibinfo {year} {2017})},\
	\Eprint {http://arxiv.org/abs/1710.05893} {arXiv:1710.05893 [astro-ph.CO]}
	\BibitemShut {NoStop}%
	\bibitem [{\citenamefont {Ezquiaga}\ and\ \citenamefont
		{Zumalacárregui}(2017)}]{Ezquiaga:2017ekz}%
	\BibitemOpen
	\bibfield  {author} {\bibinfo {author} {\bibfnamefont {J.~M.}\ \bibnamefont
			{Ezquiaga}}\ and\ \bibinfo {author} {\bibfnamefont {M.}~\bibnamefont
			{Zumalacárregui}},\ }\href {\doibase 10.1103/PhysRevLett.119.251304}
	{\bibfield  {journal} {\bibinfo  {journal} {Phys. Rev. Lett.}\ }\textbf
		{\bibinfo {volume} {119}},\ \bibinfo {pages} {251304} (\bibinfo {year}
		{2017})},\ \Eprint {http://arxiv.org/abs/1710.05901} {arXiv:1710.05901
		[astro-ph.CO]} \BibitemShut {NoStop}%
	\bibitem [{\citenamefont {Baker}\ \emph {et~al.}(2017)\citenamefont {Baker},
		\citenamefont {Bellini}, \citenamefont {Ferreira}, \citenamefont {Lagos},
		\citenamefont {Noller},\ and\ \citenamefont {Sawicki}}]{Baker:2017hug}%
	\BibitemOpen
	\bibfield  {author} {\bibinfo {author} {\bibfnamefont {T.}~\bibnamefont
			{Baker}}, \bibinfo {author} {\bibfnamefont {E.}~\bibnamefont {Bellini}},
		\bibinfo {author} {\bibfnamefont {P.~G.}\ \bibnamefont {Ferreira}}, \bibinfo
		{author} {\bibfnamefont {M.}~\bibnamefont {Lagos}}, \bibinfo {author}
		{\bibfnamefont {J.}~\bibnamefont {Noller}}, \ and\ \bibinfo {author}
		{\bibfnamefont {I.}~\bibnamefont {Sawicki}},\ }\href {\doibase
		10.1103/PhysRevLett.119.251301} {\bibfield  {journal} {\bibinfo  {journal}
			{Phys. Rev. Lett.}\ }\textbf {\bibinfo {volume} {119}},\ \bibinfo {pages}
		{251301} (\bibinfo {year} {2017})},\ \Eprint
	{http://arxiv.org/abs/1710.06394} {arXiv:1710.06394 [astro-ph.CO]}
	\BibitemShut {NoStop}%
	\bibitem [{\citenamefont {Amendola}\ \emph {et~al.}(2018)\citenamefont
		{Amendola}, \citenamefont {Kunz}, \citenamefont {Saltas},\ and\ \citenamefont
		{Sawicki}}]{Amendola:2017orw}%
	\BibitemOpen
	\bibfield  {author} {\bibinfo {author} {\bibfnamefont {L.}~\bibnamefont
			{Amendola}}, \bibinfo {author} {\bibfnamefont {M.}~\bibnamefont {Kunz}},
		\bibinfo {author} {\bibfnamefont {I.~D.}\ \bibnamefont {Saltas}}, \ and\
		\bibinfo {author} {\bibfnamefont {I.}~\bibnamefont {Sawicki}},\ }\href
	{\doibase 10.1103/PhysRevLett.120.131101} {\bibfield  {journal} {\bibinfo
			{journal} {Phys. Rev. Lett.}\ }\textbf {\bibinfo {volume} {120}},\ \bibinfo
		{pages} {131101} (\bibinfo {year} {2018})},\ \Eprint
	{http://arxiv.org/abs/1711.04825} {arXiv:1711.04825 [astro-ph.CO]}
	\BibitemShut {NoStop}%
	\bibitem [{\citenamefont {Langlois}\ \emph {et~al.}(2018)\citenamefont
		{Langlois}, \citenamefont {Saito}, \citenamefont {Yamauchi},\ and\
		\citenamefont {Noui}}]{Langlois:2017dyl}%
	\BibitemOpen
	\bibfield  {author} {\bibinfo {author} {\bibfnamefont {D.}~\bibnamefont
			{Langlois}}, \bibinfo {author} {\bibfnamefont {R.}~\bibnamefont {Saito}},
		\bibinfo {author} {\bibfnamefont {D.}~\bibnamefont {Yamauchi}}, \ and\
		\bibinfo {author} {\bibfnamefont {K.}~\bibnamefont {Noui}},\ }\href {\doibase
		10.1103/PhysRevD.97.061501} {\bibfield  {journal} {\bibinfo  {journal} {Phys.
				Rev.}\ }\textbf {\bibinfo {volume} {D97}},\ \bibinfo {pages} {061501}
		(\bibinfo {year} {2018})},\ \Eprint {http://arxiv.org/abs/1711.07403}
	{arXiv:1711.07403 [gr-qc]} \BibitemShut {NoStop}%
	\bibitem [{\citenamefont {Ezquiaga}\ and\ \citenamefont
		{Zumalacárregui}(2018)}]{Ezquiaga:2018btd}%
	\BibitemOpen
	\bibfield  {author} {\bibinfo {author} {\bibfnamefont {J.~M.}\ \bibnamefont
			{Ezquiaga}}\ and\ \bibinfo {author} {\bibfnamefont {M.}~\bibnamefont
			{Zumalacárregui}},\ }\href {\doibase 10.3389/fspas.2018.00044} {\bibfield
		{journal} {\bibinfo  {journal} {Front. Astron. Space Sci.}\ }\textbf
		{\bibinfo {volume} {5}},\ \bibinfo {pages} {44} (\bibinfo {year} {2018})},\
	\Eprint {http://arxiv.org/abs/1807.09241} {arXiv:1807.09241 [astro-ph.CO]}
	\BibitemShut {NoStop}%
	\bibitem [{\citenamefont {Motohashi}\ \emph
		{et~al.}(2018{\natexlab{a}})\citenamefont {Motohashi}, \citenamefont
		{Suyama},\ and\ \citenamefont {Yamaguchi}}]{Motohashi:2017eya}%
	\BibitemOpen
	\bibfield  {author} {\bibinfo {author} {\bibfnamefont {H.}~\bibnamefont
			{Motohashi}}, \bibinfo {author} {\bibfnamefont {T.}~\bibnamefont {Suyama}}, \
		and\ \bibinfo {author} {\bibfnamefont {M.}~\bibnamefont {Yamaguchi}},\ }\href
	{\doibase 10.7566/JPSJ.87.063401} {\bibfield  {journal} {\bibinfo  {journal}
			{J. Phys. Soc. Jap.}\ }\textbf {\bibinfo {volume} {87}},\ \bibinfo {pages}
		{063401} (\bibinfo {year} {2018}{\natexlab{a}})},\ \Eprint
	{http://arxiv.org/abs/1711.08125} {arXiv:1711.08125 [hep-th]} \BibitemShut
	{NoStop}%
	\bibitem [{\citenamefont {Motohashi}\ \emph
		{et~al.}(2018{\natexlab{b}})\citenamefont {Motohashi}, \citenamefont
		{Suyama},\ and\ \citenamefont {Yamaguchi}}]{Motohashi:2018pxg}%
	\BibitemOpen
	\bibfield  {author} {\bibinfo {author} {\bibfnamefont {H.}~\bibnamefont
			{Motohashi}}, \bibinfo {author} {\bibfnamefont {T.}~\bibnamefont {Suyama}}, \
		and\ \bibinfo {author} {\bibfnamefont {M.}~\bibnamefont {Yamaguchi}},\
	}\href@noop {} {\  (\bibinfo {year} {2018}{\natexlab{b}})},\ \Eprint
	{http://arxiv.org/abs/1804.07990} {arXiv:1804.07990 [hep-th]} \BibitemShut
	{NoStop}%
	\bibitem [{\citenamefont {Creminelli}\ \emph {et~al.}(2006)\citenamefont
		{Creminelli}, \citenamefont {Luty}, \citenamefont {Nicolis},\ and\
		\citenamefont {Senatore}}]{Creminelli:2006xe}%
	\BibitemOpen
	\bibfield  {author} {\bibinfo {author} {\bibfnamefont {P.}~\bibnamefont
			{Creminelli}}, \bibinfo {author} {\bibfnamefont {M.~A.}\ \bibnamefont
			{Luty}}, \bibinfo {author} {\bibfnamefont {A.}~\bibnamefont {Nicolis}}, \
		and\ \bibinfo {author} {\bibfnamefont {L.}~\bibnamefont {Senatore}},\ }\href
	{\doibase 10.1088/1126-6708/2006/12/080} {\bibfield  {journal} {\bibinfo
			{journal} {JHEP}\ }\textbf {\bibinfo {volume} {0612}},\ \bibinfo {pages}
		{080} (\bibinfo {year} {2006})},\ \Eprint
	{http://arxiv.org/abs/hep-th/0606090} {arXiv:hep-th/0606090 [hep-th]}
	\BibitemShut {NoStop}%
	\bibitem [{\citenamefont {Cheung}\ \emph {et~al.}(2008)\citenamefont {Cheung},
		\citenamefont {Creminelli}, \citenamefont {Fitzpatrick}, \citenamefont
		{Kaplan},\ and\ \citenamefont {Senatore}}]{Cheung:2007st}%
	\BibitemOpen
	\bibfield  {author} {\bibinfo {author} {\bibfnamefont {C.}~\bibnamefont
			{Cheung}}, \bibinfo {author} {\bibfnamefont {P.}~\bibnamefont {Creminelli}},
		\bibinfo {author} {\bibfnamefont {A.~L.}\ \bibnamefont {Fitzpatrick}},
		\bibinfo {author} {\bibfnamefont {J.}~\bibnamefont {Kaplan}}, \ and\ \bibinfo
		{author} {\bibfnamefont {L.}~\bibnamefont {Senatore}},\ }\href {\doibase
		10.1088/1126-6708/2008/03/014} {\bibfield  {journal} {\bibinfo  {journal}
			{JHEP}\ }\textbf {\bibinfo {volume} {0803}},\ \bibinfo {pages} {014}
		(\bibinfo {year} {2008})},\ \Eprint {http://arxiv.org/abs/0709.0293}
	{arXiv:0709.0293 [hep-th]} \BibitemShut {NoStop}%
	\bibitem [{\citenamefont {Horava}(2009)}]{Horava:2009uw}%
	\BibitemOpen
	\bibfield  {author} {\bibinfo {author} {\bibfnamefont {P.}~\bibnamefont
			{Horava}},\ }\href {\doibase 10.1103/PhysRevD.79.084008} {\bibfield
		{journal} {\bibinfo  {journal} {Phys.Rev.}\ }\textbf {\bibinfo {volume}
			{D79}},\ \bibinfo {pages} {084008} (\bibinfo {year} {2009})},\ \Eprint
	{http://arxiv.org/abs/0901.3775} {arXiv:0901.3775 [hep-th]} \BibitemShut
	{NoStop}%
	\bibitem [{\citenamefont {Blas}\ \emph {et~al.}(2010)\citenamefont {Blas},
		\citenamefont {Pujolas},\ and\ \citenamefont {Sibiryakov}}]{Blas:2009qj}%
	\BibitemOpen
	\bibfield  {author} {\bibinfo {author} {\bibfnamefont {D.}~\bibnamefont
			{Blas}}, \bibinfo {author} {\bibfnamefont {O.}~\bibnamefont {Pujolas}}, \
		and\ \bibinfo {author} {\bibfnamefont {S.}~\bibnamefont {Sibiryakov}},\
	}\href {\doibase 10.1103/PhysRevLett.104.181302} {\bibfield  {journal}
		{\bibinfo  {journal} {Phys.Rev.Lett.}\ }\textbf {\bibinfo {volume} {104}},\
		\bibinfo {pages} {181302} (\bibinfo {year} {2010})},\ \Eprint
	{http://arxiv.org/abs/0909.3525} {arXiv:0909.3525 [hep-th]} \BibitemShut
	{NoStop}%
	\bibitem [{\citenamefont {Gao}(2014{\natexlab{a}})}]{Gao:2014soa}%
	\BibitemOpen
	\bibfield  {author} {\bibinfo {author} {\bibfnamefont {X.}~\bibnamefont
			{Gao}},\ }\href {\doibase 10.1103/PhysRevD.90.081501} {\bibfield  {journal}
		{\bibinfo  {journal} {Phys.Rev.}\ }\textbf {\bibinfo {volume} {D90}},\
		\bibinfo {pages} {081501} (\bibinfo {year} {2014}{\natexlab{a}})},\ \Eprint
	{http://arxiv.org/abs/1406.0822} {arXiv:1406.0822 [gr-qc]} \BibitemShut
	{NoStop}%
	\bibitem [{\citenamefont {Gao}(2014{\natexlab{b}})}]{Gao:2014fra}%
	\BibitemOpen
	\bibfield  {author} {\bibinfo {author} {\bibfnamefont {X.}~\bibnamefont
			{Gao}},\ }\href {\doibase 10.1103/PhysRevD.90.104033} {\bibfield  {journal}
		{\bibinfo  {journal} {Phys.Rev.}\ }\textbf {\bibinfo {volume} {D90}},\
		\bibinfo {pages} {104033} (\bibinfo {year} {2014}{\natexlab{b}})},\ \Eprint
	{http://arxiv.org/abs/1409.6708} {arXiv:1409.6708 [gr-qc]} \BibitemShut
	{NoStop}%
	\bibitem [{\citenamefont {Fujita}\ \emph {et~al.}(2016)\citenamefont {Fujita},
		\citenamefont {Gao},\ and\ \citenamefont {Yokoyama}}]{Fujita:2015ymn}%
	\BibitemOpen
	\bibfield  {author} {\bibinfo {author} {\bibfnamefont {T.}~\bibnamefont
			{Fujita}}, \bibinfo {author} {\bibfnamefont {X.}~\bibnamefont {Gao}}, \ and\
		\bibinfo {author} {\bibfnamefont {J.}~\bibnamefont {Yokoyama}},\ }\href
	{\doibase 10.1088/1475-7516/2016/02/014} {\bibfield  {journal} {\bibinfo
			{journal} {JCAP}\ }\textbf {\bibinfo {volume} {1602}},\ \bibinfo {pages}
		{014} (\bibinfo {year} {2016})},\ \Eprint {http://arxiv.org/abs/1511.04324}
	{arXiv:1511.04324 [gr-qc]} \BibitemShut {NoStop}%
	\bibitem [{\citenamefont {Gao}\ and\ \citenamefont {Yao}(2019)}]{Gao:2018znj}%
	\BibitemOpen
	\bibfield  {author} {\bibinfo {author} {\bibfnamefont {X.}~\bibnamefont
			{Gao}}\ and\ \bibinfo {author} {\bibfnamefont {Z.-B.}\ \bibnamefont {Yao}},\
	}\href {\doibase 10.1088/1475-7516/2019/05/024} {\bibfield  {journal}
		{\bibinfo  {journal} {JCAP}\ }\textbf {\bibinfo {volume} {1905}},\ \bibinfo
		{pages} {024} (\bibinfo {year} {2019})},\ \Eprint
	{http://arxiv.org/abs/1806.02811} {arXiv:1806.02811 [gr-qc]} \BibitemShut
	{NoStop}%
	\bibitem [{\citenamefont {Gao}\ \emph {et~al.}(2019{\natexlab{a}})\citenamefont
		{Gao}, \citenamefont {Kang},\ and\ \citenamefont {Yao}}]{Gao:2019lpz}%
	\BibitemOpen
	\bibfield  {author} {\bibinfo {author} {\bibfnamefont {X.}~\bibnamefont
			{Gao}}, \bibinfo {author} {\bibfnamefont {C.}~\bibnamefont {Kang}}, \ and\
		\bibinfo {author} {\bibfnamefont {Z.-B.}\ \bibnamefont {Yao}},\ }\href
	{\doibase 10.1103/PhysRevD.99.104015} {\bibfield  {journal} {\bibinfo
			{journal} {Phys. Rev.}\ }\textbf {\bibinfo {volume} {D99}},\ \bibinfo {pages}
		{104015} (\bibinfo {year} {2019}{\natexlab{a}})},\ \Eprint
	{http://arxiv.org/abs/1902.07702} {arXiv:1902.07702 [gr-qc]} \BibitemShut
	{NoStop}%
	\bibitem [{\citenamefont {Gao}\ \emph {et~al.}(2019{\natexlab{b}})\citenamefont
		{Gao}, \citenamefont {Yamaguchi},\ and\ \citenamefont
		{Yoshida}}]{Gao:2018izs}%
	\BibitemOpen
	\bibfield  {author} {\bibinfo {author} {\bibfnamefont {X.}~\bibnamefont
			{Gao}}, \bibinfo {author} {\bibfnamefont {M.}~\bibnamefont {Yamaguchi}}, \
		and\ \bibinfo {author} {\bibfnamefont {D.}~\bibnamefont {Yoshida}},\ }\href
	{\doibase 10.1088/1475-7516/2019/03/006} {\bibfield  {journal} {\bibinfo
			{journal} {JCAP}\ }\textbf {\bibinfo {volume} {1903}},\ \bibinfo {pages}
		{006} (\bibinfo {year} {2019}{\natexlab{b}})},\ \Eprint
	{http://arxiv.org/abs/1810.07434} {arXiv:1810.07434 [hep-th]} \BibitemShut
	{NoStop}%
	\bibitem [{\citenamefont {Gao}\ and\ \citenamefont {Yao}(2020)}]{Gao:2019twq}%
	\BibitemOpen
	\bibfield  {author} {\bibinfo {author} {\bibfnamefont {X.}~\bibnamefont
			{Gao}}\ and\ \bibinfo {author} {\bibfnamefont {Z.-B.}\ \bibnamefont {Yao}},\
	}\href {\doibase 10.1103/PhysRevD.101.064018} {\bibfield  {journal} {\bibinfo
			{journal} {Phys. Rev.}\ }\textbf {\bibinfo {volume} {D101}},\ \bibinfo
		{pages} {064018} (\bibinfo {year} {2020})},\ \Eprint
	{http://arxiv.org/abs/1910.13995} {arXiv:1910.13995 [gr-qc]} \BibitemShut
	{NoStop}%
	\bibitem [{\citenamefont {Gao}\ and\ \citenamefont {Hong}(2020)}]{Gao:2019liu}%
	\BibitemOpen
	\bibfield  {author} {\bibinfo {author} {\bibfnamefont {X.}~\bibnamefont
			{Gao}}\ and\ \bibinfo {author} {\bibfnamefont {X.-Y.}\ \bibnamefont {Hong}},\
	}\href {\doibase 10.1103/PhysRevD.101.064057} {\bibfield  {journal} {\bibinfo
			{journal} {Phys.\ Rev.\ D}\ }\textbf {\bibinfo {volume} {101}},\ \bibinfo
		{pages} {064057} (\bibinfo {year} {2020})},\ \Eprint
	{http://arxiv.org/abs/1906.07131} {arXiv:1906.07131 [gr-qc]} \BibitemShut
	{NoStop}%
	\bibitem [{\citenamefont {Jacobson}\ and\ \citenamefont
		{Mattingly}(2001)}]{Jacobson:2000xp}%
	\BibitemOpen
	\bibfield  {author} {\bibinfo {author} {\bibfnamefont {T.}~\bibnamefont
			{Jacobson}}\ and\ \bibinfo {author} {\bibfnamefont {D.}~\bibnamefont
			{Mattingly}},\ }\href {\doibase 10.1103/PhysRevD.64.024028} {\bibfield
		{journal} {\bibinfo  {journal} {Phys.Rev.}\ }\textbf {\bibinfo {volume}
			{D64}},\ \bibinfo {pages} {024028} (\bibinfo {year} {2001})},\ \Eprint
	{http://arxiv.org/abs/gr-qc/0007031} {arXiv:gr-qc/0007031 [gr-qc]}
	\BibitemShut {NoStop}%
	\bibitem [{\citenamefont {Gleyzes}\ \emph {et~al.}(2013)\citenamefont
		{Gleyzes}, \citenamefont {Langlois}, \citenamefont {Piazza},\ and\
		\citenamefont {Vernizzi}}]{Gleyzes:2013ooa}%
	\BibitemOpen
	\bibfield  {author} {\bibinfo {author} {\bibfnamefont {J.}~\bibnamefont
			{Gleyzes}}, \bibinfo {author} {\bibfnamefont {D.}~\bibnamefont {Langlois}},
		\bibinfo {author} {\bibfnamefont {F.}~\bibnamefont {Piazza}}, \ and\ \bibinfo
		{author} {\bibfnamefont {F.}~\bibnamefont {Vernizzi}},\ }\href {\doibase
		10.1088/1475-7516/2013/08/025} {\bibfield  {journal} {\bibinfo  {journal}
			{JCAP}\ }\textbf {\bibinfo {volume} {1308}},\ \bibinfo {pages} {025}
		(\bibinfo {year} {2013})},\ \Eprint {http://arxiv.org/abs/1304.4840}
	{arXiv:1304.4840 [hep-th]} \BibitemShut {NoStop}%
	\bibitem [{\citenamefont {Germani}\ \emph {et~al.}(2009)\citenamefont
		{Germani}, \citenamefont {Kehagias},\ and\ \citenamefont
		{Sfetsos}}]{Germani:2009yt}%
	\BibitemOpen
	\bibfield  {author} {\bibinfo {author} {\bibfnamefont {C.}~\bibnamefont
			{Germani}}, \bibinfo {author} {\bibfnamefont {A.}~\bibnamefont {Kehagias}}, \
		and\ \bibinfo {author} {\bibfnamefont {K.}~\bibnamefont {Sfetsos}},\ }\href
	{\doibase 10.1088/1126-6708/2009/09/060} {\bibfield  {journal} {\bibinfo
			{journal} {JHEP}\ }\textbf {\bibinfo {volume} {0909}},\ \bibinfo {pages}
		{060} (\bibinfo {year} {2009})},\ \Eprint {http://arxiv.org/abs/0906.1201}
	{arXiv:0906.1201 [hep-th]} \BibitemShut {NoStop}%
	\bibitem [{\citenamefont {Blas}\ \emph {et~al.}(2009)\citenamefont {Blas},
		\citenamefont {Pujolas},\ and\ \citenamefont {Sibiryakov}}]{Blas:2009yd}%
	\BibitemOpen
	\bibfield  {author} {\bibinfo {author} {\bibfnamefont {D.}~\bibnamefont
			{Blas}}, \bibinfo {author} {\bibfnamefont {O.}~\bibnamefont {Pujolas}}, \
		and\ \bibinfo {author} {\bibfnamefont {S.}~\bibnamefont {Sibiryakov}},\
	}\href {\doibase 10.1088/1126-6708/2009/10/029} {\bibfield  {journal}
		{\bibinfo  {journal} {JHEP}\ }\textbf {\bibinfo {volume} {0910}},\ \bibinfo
		{pages} {029} (\bibinfo {year} {2009})},\ \Eprint
	{http://arxiv.org/abs/0906.3046} {arXiv:0906.3046 [hep-th]} \BibitemShut
	{NoStop}%
	\bibitem [{\citenamefont {Jacobson}(2010)}]{Jacobson:2010mx}%
	\BibitemOpen
	\bibfield  {author} {\bibinfo {author} {\bibfnamefont {T.}~\bibnamefont
			{Jacobson}},\ }\href {\doibase 10.1103/PhysRevD.81.101502} {\bibfield
		{journal} {\bibinfo  {journal} {Phys.\ Rev.\ D}\ }\textbf {\bibinfo {volume}
			{81}},\ \bibinfo {pages} {101502} (\bibinfo {year} {2010})},\ \bibinfo {note}
	{[Erratum: Phys.Rev.D 82, 129901 (2010)]},\ \Eprint
	{http://arxiv.org/abs/1001.4823} {arXiv:1001.4823 [hep-th]} \BibitemShut
	{NoStop}%
	\bibitem [{\citenamefont {Blas}\ \emph {et~al.}(2011)\citenamefont {Blas},
		\citenamefont {Pujolas},\ and\ \citenamefont {Sibiryakov}}]{Blas:2010hb}%
	\BibitemOpen
	\bibfield  {author} {\bibinfo {author} {\bibfnamefont {D.}~\bibnamefont
			{Blas}}, \bibinfo {author} {\bibfnamefont {O.}~\bibnamefont {Pujolas}}, \
		and\ \bibinfo {author} {\bibfnamefont {S.}~\bibnamefont {Sibiryakov}},\
	}\href {\doibase 10.1007/JHEP04(2011)018} {\bibfield  {journal} {\bibinfo
			{journal} {JHEP}\ }\textbf {\bibinfo {volume} {1104}},\ \bibinfo {pages}
		{018} (\bibinfo {year} {2011})},\ \Eprint {http://arxiv.org/abs/1007.3503}
	{arXiv:1007.3503 [hep-th]} \BibitemShut {NoStop}%
	\bibitem [{\citenamefont {Chagoya}\ and\ \citenamefont
		{Tasinato}(2018)}]{Chagoya:2018yna}%
	\BibitemOpen
	\bibfield  {author} {\bibinfo {author} {\bibfnamefont {J.}~\bibnamefont
			{Chagoya}}\ and\ \bibinfo {author} {\bibfnamefont {G.}~\bibnamefont
			{Tasinato}},\ }\href@noop {} {\  (\bibinfo {year} {2018})},\ \Eprint
	{http://arxiv.org/abs/1805.12010} {arXiv:1805.12010 [hep-th]} \BibitemShut
	{NoStop}%
	\bibitem [{\citenamefont {Zhu}\ \emph {et~al.}(2012)\citenamefont {Zhu},
		\citenamefont {Shu}, \citenamefont {Wu},\ and\ \citenamefont
		{Wang}}]{Zhu:2011yu}%
	\BibitemOpen
	\bibfield  {author} {\bibinfo {author} {\bibfnamefont {T.}~\bibnamefont
			{Zhu}}, \bibinfo {author} {\bibfnamefont {F.-W.}\ \bibnamefont {Shu}},
		\bibinfo {author} {\bibfnamefont {Q.}~\bibnamefont {Wu}}, \ and\ \bibinfo
		{author} {\bibfnamefont {A.}~\bibnamefont {Wang}},\ }\href {\doibase
		10.1103/PhysRevD.85.044053} {\bibfield  {journal} {\bibinfo  {journal} {Phys.
				Rev.}\ }\textbf {\bibinfo {volume} {D85}},\ \bibinfo {pages} {044053}
		(\bibinfo {year} {2012})},\ \Eprint {http://arxiv.org/abs/1110.5106}
	{arXiv:1110.5106 [hep-th]} \BibitemShut {NoStop}%
	\bibitem [{\citenamefont {Dvali}\ \emph {et~al.}(2000)\citenamefont {Dvali},
		\citenamefont {Gabadadze},\ and\ \citenamefont {Porrati}}]{Dvali:2000hr}%
	\BibitemOpen
	\bibfield  {author} {\bibinfo {author} {\bibfnamefont {G.}~\bibnamefont
			{Dvali}}, \bibinfo {author} {\bibfnamefont {G.}~\bibnamefont {Gabadadze}}, \
		and\ \bibinfo {author} {\bibfnamefont {M.}~\bibnamefont {Porrati}},\ }\href
	{\doibase 10.1016/S0370-2693(00)00669-9} {\bibfield  {journal} {\bibinfo
			{journal} {Phys.Lett.}\ }\textbf {\bibinfo {volume} {B485}},\ \bibinfo
		{pages} {208} (\bibinfo {year} {2000})},\ \Eprint
	{http://arxiv.org/abs/hep-th/0005016} {arXiv:hep-th/0005016 [hep-th]}
	\BibitemShut {NoStop}%
	\bibitem [{\citenamefont {Deffayet}\ \emph {et~al.}(2015)\citenamefont
		{Deffayet}, \citenamefont {Esposito-Farese},\ and\ \citenamefont
		{Steer}}]{Deffayet:2015qwa}%
	\BibitemOpen
	\bibfield  {author} {\bibinfo {author} {\bibfnamefont {C.}~\bibnamefont
			{Deffayet}}, \bibinfo {author} {\bibfnamefont {G.}~\bibnamefont
			{Esposito-Farese}}, \ and\ \bibinfo {author} {\bibfnamefont {D.~A.}\
			\bibnamefont {Steer}},\ }\href {\doibase 10.1103/PhysRevD.92.084013}
	{\bibfield  {journal} {\bibinfo  {journal} {Phys. Rev.}\ }\textbf {\bibinfo
			{volume} {D92}},\ \bibinfo {pages} {084013} (\bibinfo {year} {2015})},\
	\Eprint {http://arxiv.org/abs/1506.01974} {arXiv:1506.01974 [gr-qc]}
	\BibitemShut {NoStop}%
	\bibitem [{\citenamefont {Gao}()}]{Gao:inprogress}%
	\BibitemOpen
	\bibfield  {author} {\bibinfo {author} {\bibfnamefont {X.}~\bibnamefont
			{Gao}},\ }\href@noop {} {\bibinfo  {journal} {work in progress}\
	}\BibitemShut {NoStop}%
	\bibitem [{\citenamefont {De~Felice}\ \emph {et~al.}(2018)\citenamefont
		{De~Felice}, \citenamefont {Langlois}, \citenamefont {Mukohyama},
		\citenamefont {Noui},\ and\ \citenamefont {Wang}}]{DeFelice:2018ewo}%
	\BibitemOpen
	\bibfield  {journal} {  }\bibfield  {author} {\bibinfo {author} {\bibfnamefont
			{A.}~\bibnamefont {De~Felice}}, \bibinfo {author} {\bibfnamefont
			{D.}~\bibnamefont {Langlois}}, \bibinfo {author} {\bibfnamefont
			{S.}~\bibnamefont {Mukohyama}}, \bibinfo {author} {\bibfnamefont
			{K.}~\bibnamefont {Noui}}, \ and\ \bibinfo {author} {\bibfnamefont
			{A.}~\bibnamefont {Wang}},\ }\href {\doibase 10.1103/PhysRevD.98.084024}
	{\bibfield  {journal} {\bibinfo  {journal} {Phys. Rev. D}\ }\textbf {\bibinfo
			{volume} {98}},\ \bibinfo {pages} {084024} (\bibinfo {year} {2018})},\
	\Eprint {http://arxiv.org/abs/1803.06241} {arXiv:1803.06241 [hep-th]}
	\BibitemShut {NoStop}%
\end{thebibliography}
\end{document}